# F.W. Longbottom: astronomical photographer and founder of the Chester Astronomical Society

**Jeremy Shears**

## Abstract

Frederick William Longbottom FRAS (1850-1933) was an original member of the BAA and served as Director of its Photographic Section between 1906 and 1926. A hop merchant by trade, he spent much of his life in Chester where he was instrumental in founding the City's first astronomical society in 1892.

## Introduction

In his book "*The Victorian Amateur Astronomer*" (1), Dr Allan Chapman describes how during the second half of the nineteenth century a profound social change took place in British astronomy. Initially the subject was dominated by the independently wealthy "Grand Amateurs". Later in the century the comfortable and educated middle classes, with leisure time on their hands, began to take up various hobbies including astronomy. Around the country amateur astronomical societies began to be formed which allowed people to discuss their shared interest, listen to lectures on the subject and learn how and what to observe with their telescopes. Equally important was the social aspect, allowing like-minded men and women to enjoy each others' company, no matter how committed they were to the subject. In many ways this is similar to the dynamic in the myriad of local astronomical societies that exist across the country today.

The oldest of the local societies is Leeds Astronomical Society. Although this was formed in 1859 it had a couple of false starts (2), eventually taking off in 1892. Next came the Liverpool Astronomical Society (LAS) in 1881 which for a period was a society of national standing with members from all over the UK and even overseas. The BAA was formed in 1890 and soon spawned local branches. A North-West Branch (1892 (3)), drawing members from Manchester and neighbouring parts of Cheshire, was short lived and eventually sought independence as the new Manchester Astronomical Society. Branches soon followed in Glasgow (1894) (4), Edinburgh (1896) (5) and Birmingham (1900) (6). Astronomical societies were formed in a similar period in Newcastle-upon-Tyne (1904) and Cardiff.

It is perhaps not surprising that the new astronomical societies were associated with the major industrial centres, but what about the provincial cities and towns? In 1892, Chester became one of the earliest of the smaller cities to have its own astronomical society. Frederick William Longbottom (1850-1933; Figure 1) was instrumental in the founding of this society. He was also an original member of the BAA and he went on to direct the BAA Photographic Section for 20 years. This paper traces his involvement in these activities and discusses his own astronomical pursuits.

## Family and career

Frederick William Longbottom was born at Scarborough, Yorkshire, on 8 August 1850. His father was Joseph Longbottom (1819-1889) and his mother Sarah Longbottom (7). Frederick, or Fred as he was widely known throughout his life, was the youngest of 4 children (8). In 1861 Joseph was earning a living dealing in antique china in Leeds, but by the 1880s the family had moved to Worcester where Joseph became a partner in the hop merchants, Piercy, Longbottom & Faram. Worcestershire was, and still is, one of the main hop-growing regions of the country. In 1885 Charles Faram went his separate way, forming Charles Faram & Co. Ltd (9), which still trades in the Malvern Hills in Worcestershire. Joseph Longbottom and George Piercy continued as partners in their own hop merchants business in Worcester and expanded into sales of seeds (10).

Frederick Longbottom joined his father in the family company and eventually a second office was opened in Manchester. Frederick and Joseph set up residence in Chester in the late 1880s living in Queen’s Park and Curzon Park respectively (11). These were two of the most affluent and fashionable parts of the City, close to the River Dee, which had been developed following the City's growing prosperity in the mid 19th century. When Joseph died in 1889, Frederick took on his father’s share of the business, but he eventually sold out to his partner, George Piercy, in 1894. Sometime after this Frederick retired (12), allowing him more time to indulge his hobbies, about which we will hear more later.

Frederick was married to Emily Anne Longbottom, 5 years his junior, and they had one child, Frederick Hindle Cecil Longbottom, who died in infancy (13). They remained in Chester until 1921 (3), after which they moved about for several years, with addresses variously in Malvern, Worcestershire, and London (14), before finally setting up home in Boscombe, Hampshire (15) in 1925. Frederick Longbottom died in Boscombe on 7 March 1933 and Emily died on 7 January 1945 (16).

## A clubbable interest in astronomy

Longbottom’s interest in astronomy is said to have originated in boyhood (17) during a school holiday which he “spent with a taciturn uncle, so that he had to read all the evenings, one book happened to be about astronomy and that started his life-long interest in the science” (18). He was sufficiently committed to the hobby to join the LAS in 1885 (19), which as noted before, drew a national membership, whilst still living in Worcestershire. The following year he wrote the first of many letters to the English Mechanic on astronomical matters (20), in which he discussed the advantages of metal versus wooden tubes for Newtonian reflectors and mentioned the excellent quality of his 6½ inch mirror which had been made by G.H. With (1827-1904) of Hereford.

When the idea of a national association for amateur astronomers was suggested by W.H.S. Monck in the summer of 1890, following the implosion of the LAS,

Longbottom was enthusiastic about the prospect and immediately joined, thus becoming one of original members of what was to become the BAA. He was present at the first general meeting in London on Friday 24 October 1890 (21). The meeting was largely given over to organisational and administrative aspects of the fledgling society and one important item that needed to be agreed was the starting time of its main meetings. Longbottom spoke in favour of a later starting time of 7 pm. Results were presented from a postal ballot of members which was not conclusive and a vote at the meeting resulted in a starting time of 5 pm being adopted.

Longbottom attended many BAA meetings over the years, often combining them with business travel, and spoke frequently. He become well-known known for his “bright manner and merry quips” (17) and his “genial and generous disposition” (18). He often injected some humour into the proceedings. During the June 1892 meeting Albert Marth (1828-1897) retold the story, almost certainly untrue, of how Challis might have missed discovering Neptune by having a cup of tea. Challis was entertaining friends at his house and on noticing the sky clearing made to go out to the observatory and find the reported object. But the hostess was heard to say "I cannot let you go till you have had your tea, which will be ready in a few minutes". Half an hour later the clear sky had disappeared and the opportunity was lost. In the questions which followed, Longbottom quipped “that the incident of the tea-party pointed to the conclusion that ladies should be made astronomers as soon as possible”. (22)

Longbottom was elected Fellow of the Royal Astronomical Society in 1905, having been proposed by his good friend, W.E. Plummer (1849-1928). Plummer (Figure 2) had been Director of the Liverpool Observatory at Bidston Hill, near Birkenhead, since 1892 and was personally well-acquainted with Longbottom though their membership of the LAS and, as well shall see later, the Chester Astronomical Society. Longbottom was also a member of the Astronomical Society of Wales which had been formed by Arthur Mee (1860-1926) in 1894 and, despite its name, also attracted some of the more active astronomers from outside the Principality (23).

Over the years Longbottom owned many telescopes and one wonders whether he might sometimes have obtained as much enjoyment from owning and testing them, as he did from actually using them. Over the years he discussed the relative performance of different telescopes in the *English Mechanic*. The list of instruments contains the names of many well-known telescope manufacturers, including a 3¼ inch Wray refractor (1886), a 6½ inch With reflector (1887), a 3½ inch Bardou refractor (1900), a 10¼ inch Newtonian (1905), a 15 inch Cope reflector, a 3 inch Broadhurst Clarkson refractor (1913), a 6 inch Grubb refractor (1915), a 12½ inch Cope reflector (1920), a 6½ inch reflector with a mirror by With (1920), a 6 inch Cooke refractor (1927), a 4 inch Cooke refractor (1933), and 3 inch wide field “comet seeker” refractor (1933). His largest telescope was an 18½ inch reflector with a Calver mirror which he obtained shortly after moving to Chester.

**Astronomical photography and the BAA Photographic Section**

Longbottom was an active observer of the moon, planets and comets (24), but from the mid-1890’s he became increasingly interested in astronomical photography. Much of his early work was conducted with his 18½ inch Calver, an exceptionally large amateur instrument for the period. This had the advantage of considerable light grasp, as well as being on a mount which “was as steady as a rock – it took half a gale to move that!” (25). On the other hand it did have two significant drawbacks. The first was that the mount was an altazimuth, which meant it was only suitable for short exposures of bright objects such as the moon and planets. Secondly, Longbottom had a restricted view of the sky from his Chester garden and obviously the telescope was too massive to move around the garden, which is why he sometimes used smaller and more portable telescopes (26). An 1896 photograph of the Moon is shown in Figure 3. The following year he undertook a series of photographs of Venus and Jupiter, noting in a letter to the English Mechanic (27):

“Jupiter came out well … and the negative shows about as much detail as was visible to the eye in the telescope about five minutes afterwards, but no satellites. I am writing this letter to ask others to co-operate with me in photographing Venus during next few evenings, as I fancy my pictures suggest the short period of rotation, the N. cusp being more blunted on some negatives than on others. With 18½ inches aperture on a Mawson lantern plate, and an exposure of 1/60 sec, long development brings up a fairly dense image of this planet”.

Figure 4 shows three images of the crescent Venus captured on the same plate with the 18½ inch telescope. The images appear to have been taken at the prime focus using a Barlow lens of unspecified power. As such the planet is only 1 to 2 mm in diameter. Quite what lead him to believe he might have detected rotation is not clear, especially given the small size of the images.

Further experiments with meteor (28) and comet photography followed and very soon he began to earn a reputation as a competent astrophotographer. In later years he was invited to contribute a section on “Celestial Photography” in T.E.R. Phillips (1868-1942) and W.H. Steavenson’s (1894-1975) classic *Splendour of the Heavens* (1923) (29) and the Preface to H.H. Waters’ *Astronomical Photography for Amateurs* (1921) (30), one of the first texts dedicated to this subject.

The BAA initially formed a Photographic Section in December 1891 under the Directorship of William Schooling (1860-1936), with the aim of encouraging members to adopt this fast developing technique, as well as producing a collection of lantern slides for use by the Association to illustrate its meetings. Unfortunately, the scarcity of suitable apparatus amongst members seriously restricted the activities of the Section (31) and the Director disbanded it within a year. Nevertheless, the idea of a Photographic Section was a good one, perhaps just a little ahead of its time. This came in 1897 when the Section was re-established under Joseph Lunt (1866-1940).

Lunt set about implementing “a preconceived plan of campaign which included solar, lunar and stellar photography, and the photographic investigation of meteors, comets, new and variable stars, nebulae and clusters” (31). However, within a few months Lunt accepted an appointment as Assistant at the Royal Observatory at the Cape of Good Hope and tendered his resignation (32). In 1898 R. Wilding was appointed to succeed Lunt as Director, and with his increasing profile as an astrophotographer, it was not surprising that Longbottom was asked to be Assistant Director.

By 1902 the Photographic Section had 37 members and the membership list identified Longbottom as using a 3 inch refractor and 2½ and 5½ inch portrait lenses, as well as the 18½ inch Calver (33). In 1906 Wilding was unable to retain the Directorship due to his frequent absences from England and Longbottom was appointed by Council to succeed him. This was announced at the November 1906 meeting by BAA President F.W. Levander, FRAS, allowing Longbottom to respond in his typical self-deprecating style as recorded in the meeting minutes: (34)

“There was only one qualification that he [Longbottom] could bring to bear upon the Section that might be more favourable than in the case of Mr. Wilding—he was pretty constantly at home. In place of great ability he must substitute great enthusiasm”.

Longbottom went on to address the meeting about the subject in which he was becoming increasingly active: comet photography. He pointed out that “Prof. Barnard [E.E. Barnard (1857-1923)] had said he thought every comet ought to be photographed, if possible, every hour from its first appearance, and the life history of a comet taken in that way would certainly be of the deepest interest, and would throw important light on cometary matters”. Comet Section Director, Dr. A.C. D. Crommelin (1865-1939) had already drawn attention to Longbottom’s photographs of comets in 1897 (35) and by 1905 his long exposures, often 90 mins in duration and all hand guided, were being regularly shown at the BAA meetings, receiving praise from astronomers such as E.W. Maunder (1851-1928).

One of the brighter comets of the early 1900s was Comet Morehouse 1908e (C/1908 R1). The comet was discovered on 1 September 1908 and Longbottom had his first view of it on 12 September when he swept it up with his 3½ inch binocular (36). He followed it for the next few nights with his 6 inch Grubb refractor, noting changes in the comet’s tail, and it finally became visible to the naked eye on 29 September. He obtained 34 photographs of it between September 15 and October 25, with exposures of 3 to 53 mins. For the majority of these he used a purpose-built 12½ inch reflector with an unusually short focal length of 24 inches (f1.9) which he had had specially constructed for photography by Cope of West Malvern, Worcestershire. Some examples of Longbottom’s photographs of the comet are shown in Figure 5; the one of 15 October shows a major tail disconnection event, which he described as follows:

“there was a most remarkable “elbow” in the principal tail. Commencing about half a degree from the nucleus, the cometary matter turned at an angle of 45°, and continued at this for quite 15 minutes, when it turned sharply almost parallel with the main axis. At the curve the brightness was much greater, as if the material flowing from the head accumulated in trying to escape round the bend. This bright area looked in the telescope almost like a second comet; but the photographs show the curved tail to be quite continuous. Running from the head was a fainter straight tail. Clouds prevented long exposures; but two plates were secured with exposures of 10 and 5 minutes, half-an-hour apart, and these prove that the bent portion of the tail was in rapid motion— wagging perhaps!”

We now know that that such disconnection events are in fact caused by an interaction between particles in the comet’s ion tail and the magnetic field entrained in a coronal mass ejection (CME) event carried on the solar wind.

Longbottom also took a long series of photographs of another bright comet, Comet Brooks 1911c (C/1911 O1), which showed how its tail evolved. He took 56 plates of the comet between 25 July and 1 October 1911 using his 15-inch f/4 Cope reflector and a Dallmeyer 3-in. f/2.7 portrait lens (37). Two photographs of the comet showing changes in the tail between 21 and 24 September are shown in Figure 6. Longbottom's photographs of Comet Brooks came in for some criticism for possible guiding errors in a subsequent edition of the Journal. Both R.A. Sampson (1866-1939) and N. Maclachlan wrote letters (38) commenting on the unusual trailing of the background stars, even suggesting that Longbottom might have mistaken the date on which the plates were taken. Longbottom replied, stating the dates were correct and admitting that his guiding was not perfect and he challenged his critics to share their efforts with readers (39).

Longbottom photographed many other comets, especially in the years leading up to the First World War. Although the photographs were frequently shown at BAA meetings, relatively few appear in print. Upon his death, he donated his photographic collection to the BAA, but its present whereabouts are not known and are feared lost (40).

During the war, the activities of the Photographic Section were at a low ebb as Longbottom noted in his Director’s report for 1915/1916 (41):

“Work in the Section has almost come to a standstill…Observers find their energies have to be given to more urgent matters”.

There was an upside, though, for those who still had time for astronomy: “The lighting restrictions have much improved conditions for long exposure photography, the background of the sky being much darker”

Moreover, it became increasingly difficult to obtain the necessary equipment, especially photographic plates, although Longbottom was able to continue his work throughout the war. Even when peace returned it took several years for supplies of plates and chemicals for their development to become reliable. Consequently the work of the photographic section took some time to recover. On the other hand, G.F. Kellaway (1902-1962), a later Section Director (1937 – 1950), noted: (42) “The effects of the War of 1914-1918 had not been entirely adverse to the Section’s activities. The belligerents on both sides had been constantly improving their method of aerial photography and in their efforts to combine a large image with a short exposure had produced large fast anastigmatic lenses with a wide covering power, admirably suited to the more peaceful pursuit of celestial photography”.

One specific project that was thwarted by the war had been proposed by section member Lawrence Richardson (1869-1953), of Newcastle-upon-Tyne, to produce a “Photographic Star Atlas to the Ninth Magnitude”. Longbottom announced the project at the November 1914 meeting of the BAA. A prerequisite was that all contributors should use identical lenses to ensure the photographs were of the same scale. This would have been an organisational and financial challenge at the best of times, given that an order would have to be made to a manufacturer for a batch of such lenses, but with the onset of war the difficulties were insurmountable. Even though the project never really took off as a collaborative effort, Richardson continued the project largely on his own (43). Photographs by Longbottom, Richardson, Rev. Walter Bidlake (44), of Crewe, and 4 other amateur astronomers were used by Rev. T.H.E.C. Espin (1858-1934) in his “*Survey of dark structures in the milky way*” published in 1922 (45), the aim of which was to identify and catalogue dark nebulae which had not been included in E.E. Barnard’s catalogue (46)of these objects.

Longbottom was unable to do much photography during his itinerant years after leaving Chester and before settling down permanently in Boscombe, although he kept up the Section’s correspondence. In 1926 he announced (47) his retirement as Director having held the office for 20 years. His successor was F.J. Hargreaves (1891-1970). After this point, Longbottom focussed on wide field photography of meteors as well as visual observations, especially of Venus (48).

## Eclipse expeditions

During the first decade and a half of the existence of the BAA, expeditions were organised to four total solar eclipses, in 1896, 1898, 1900 and 1905. These were not solely scientific enterprises, although science was the aim of many. Others joined the expeditions out of general interest with a desire to marvel at one of nature’s wonders, some simply wanted to enjoy the excitement of travel to foreign climes and to enjoy the company of like-minded people. Longbottom joined three of these eclipse expeditions, only missing the 1898 expedition to India.

### *Norway 1896*

Longbottom’s first expedition was to Norway for the eclipse of 9 August 1896, which saw a contingent of 58 BAA members set sail from Tilbury on 25 July on board the *Norse King*. Meticulous plans were laid down and discussed with observers, and detailed rehearsals at the site near Vadső, but in the event the party was clouded out (49). Longbottom had taken a 6 inch reflector with which he had intended to photograph the eclipse. Despite the cloud, Longbottom noted: (50)

"In the early morning of 'the great day' the party stood on the summit of a hill, breathless and excited. Suddenly, as if some invisible hand had swept across the mountain, they were enveloped in a total shadow. They could not liken it to night, but to a funeral pall with a fringe of gold".

*Algiers 1900*

The 1900 eclipse in Algiers on 28 May was much more successful. It was originally intended for the BAA to charter a ship, but these plans had to be abandoned with the onset of the Boer War in South Africa which meant that the shipping lines could not guarantee availability of ships as many were required to transport troops (51). Thus it was up to individual members, or groups of members, to make their own arrangements. Longbottom elected to go to Algiers on the Orient Steamship Company's steam yacht *Argonaut* (52). The party included Longbottom's good friend, Dr. Harold Whichello (1870-1945). Whichello, a member of the BAA and the LAS, was a GP in the village of Tattenhall 14 km south-east of Chester (53).

The *Argonaut* arrived in Algiers on the morning of 27 May. Longbottom and Whichello's party of about 40 people disembarked at Cape Mantifou, the north-east horn of the Bay of Algiers. Whichello and Harry Krauss Nield (54), another Cheshire amateur astronomer, were despatched to locate a suitable observing location for the party. Krauss Nield wrote: (51)

"The first likely place we saw was the village washing shed, and Dr. Whichello and I went, much to the embarrassment of the inmates, to survey this, but although suitable in almost every other way, the front was at rather too great an angle to the direction which would be required. After this we noticed the village school, the playground of which seemed to contain all that we desired, and we at once started making enquiries. Dr. Whichello's French being vastly superior to mine, he acted as spokesman. He first of all asked some little children if the schoolmaster was in: 'No'; 'When will he be in?' 'Never.' 'What do you mean?' 'There is no schoolmaster.' 'Who is in, then?' 'The schoolmistress.' After this we found the schoolmistress, and an obliging lady she proved to be. She said that we could use the playground and veranda of the school with the greatest of pleasure, and that she would send the children home early, so that they should be out of our way".

A photograph of some of the observers preparing to observe the eclipse at the school is shown in Figure 7. During totality Longbottom obtained some wide field photographs showing the corona, whilst other members of the party sketched the corona, including Whichello and Krauss Nield (Figure 8).

*Spain 1905*

Success was also realised during the eclipse of 30 August 1905. BAA members observed totality from a variety of locations including North America and Spain (55). Longbottom and several other members observed it at sea on board P&O's *SS Arcadia*, off the Spanish coast. Arrangements had been made by G.F. Chambers (1841-1915) with P&O for the vessel to be located on the line of totality and for the deck above the ship's smoking saloon, from where they observed the eclipse, to be whitewashed to facilitate the detection of shadow bands, which were indeed seen. Other members present included Charles Lewis Brook (1855-1939), who would later become Director of the BAA Variable Star Section and his sister Ruth Mary Brook (1856-1932), a future BAA Council member (56). Longbottom obtained a series of photographs of the eclipse (57).

*England 1927*

Longbottom's final experience of a total solar eclipse was the event on 29 June 1927. This time he didn't have to travel so far, as the path of totality crossed the UK. Although many BAA members, including a large party at Giggleswick, were disappointed, Longbottom had a decent view, noting (58) some "full pink" prominences, which were in contrast to those he saw in 1905 which were full red.

## Egypt and secondment to the Helwân Observatory

In late October 1911 Longbottom set sail for Egypt on board P&O's *SS Mongolia*, bound for Egypt, where he planned to overwinter for 4½ months. He described this as "*An Astronomical Holiday*" (59), but in practise he had made arrangements to be seconded as a volunteer to the Khedivial Observatory. The original Observatory was built in 1865 by Isma'il Pasha (1830-1895), the Ottoman Khedive (Viceroy) of Egypt and Sudan, at Abbasiya, north-east of Cairo. The Observatory moved to a new site in an area of open dessert near Helwân, 30 km south of Cairo, in 1904 to escape urban encroachment. In 1907 a new 30 inch reflector was unveiled (Figure 9), the main purpose of which was to photograph and classify nebulae with southerly declinations of between 0° and -40°, thus not visible from the UK. The telescope construction was financed by the wealthy Birmingham businessman John Henry Reynolds (1874-1949), who served as the President of the Royal Astronomical Society between 1935 and 1937. He also donated the 30 inch mirror which had been made by Dr. Andrew Ainslie Common (1841-1903) (60). The instrument was presented to the Egyptian government and named the "Reynolds Telescope".

The Observatory, under the control of the Egyptian Survey Department, was frequently short of skilled manpower with which to operate the telescope due to the many other responsibilities of the staff, hence Longbottom's assistance was greatly appreciated (61). He participated in the Observatory's photographic programme, where his experience in astronomical photography was especially helpful, exposing and developing plates and preparing them for publication (62). He worked alongside

Harold Knox-Shaw (1885-1970; Figure 10), who had joined the staff of Helwân Observatory in 1908 and would become its director in 1913 (63).

In later years Longbottom often referred fondly to his time at Helwân and he developed a continuing friendship with both Knox-Shaw and Reynolds (64). He donated a direct vision spectroscope to the Observatory (65).

The following winter, 1912/13, Longbottom travelled to Ceylon (Sri Lanka), but there is no record of whether he made any astronomical observations there.

**An interest in local history and archaeology**

Astronomy was not the only interest to make a claim on Longbottom's time. Given the rich archaeology of Chester, not least because it was located on the site of the Roman garrison town of *Deva*, it was perhaps not surprising that he indulged a passion for archaeology and local history. He was a member of the Chester & North Wales Archaeological Society  and in 1905 he conducted a search of the north Wirral shore near the important archaeological site of Meols with Robert Newstead (1859 – 1947: Figure 11) (66). Newstead had become curator of Chester's newly opened Grosvenor Museum in 1886 and from then on increased his involvement with Chester archaeology as well as pursuing a study of entomology. In 1905, Newstead was appointed as lecturer in Entymology and Parasitology at the Liverpool University School of Tropical Medicine, where he did pioneering work on the Tse Tse fly and its transmission of sleeping sickness, becoming professor in 1911. Nevertheless he continued his archaeological excavations, unearthing many of the major Roman sites in Chester including the amphitheatre. Although Longbottom and Newstead only found a few artefacts during their search at Meols, Longbottom wrote a paper on an earlier find of coins at the coastal site, which are on display at the Grosvenor Museum (67).

Longbottom was himself a keen numismatist and he donated several coins from his personal collection to the Grosvenor Museum, including two from the reign of Edward I (68), which were supposed to have been minted at Chester. After his death his collection of  "English Coins Including the Important Series of Halfpence and Farthings of England, Scotland and Ireland; also Greek and Roman Coins" was sold at auction at Sotheby's in May 1934 and has been described a one of the top 50 most important auctions of British coins between 1802 and 1977 (69). He also published a catalogue of Roman coins held in the Grosvenor Museum (70), a history of Church heraldry in Cheshire and, at the request of Chester council, an account of Chester during the First World War (71).

Longbottom did not neglect the history of astronomy either. Stimulated by the setting up of the BAA Historical Section in 1930, he donated his collection of "Autograph Letters and Portraits of Astronomers" which he hoped would act as a nucleus of historical documents that might be added to over the years (72). Initially the material was on view in the BAA Library. Sadly, as with his collection of astronomical photographs, no records exist of where these items are now and they too are feared lost.

**An Astronomical Society for Chester?**

Chester has a proud astronomical heritage which includes John Wilkins (1614-1672), Bishop of Chester (1668-1672), a founding member of the Royal Society and author of the popular astronomy works, *The Discovery of a World in the Moone* (1638) and *A Discourse Concerning a New Planet* (1640). Edmond Halley (1656-1742) was also controller of the Mint at Chester in 1696-7. However, not long after his arrival in Chester, Longbottom noticed that something was missing from its contemporary cultural scene: there was no astronomical society. Writing to the *Cheshire Observer* in November he commented: (73)

“Perhaps as a comparative stranger in Chester, I ought to apologise for…publicly advocating a study [i.e. astronomy] which has been, no doubt, often extolled by older and wiser men in your midst; but my earnest desire to foster this, at once the oldest, and even the youngest, of the sciences, must be my justification. Just a wee sensation of disappointment shadowed me when I came to live in a city famous for its culture, and found astronomy claiming only a straggling and disconnected adherence”.

Of course, Longbottom had been a member of the LAS since 1885 and it was a relatively brief rail journey from Chester to Liverpool to attend meetings (in 1901 Longbottom went on to serve as vice president of the LAS, at the same time that his friend W.E. Plummer was president). But the LAS had been through difficult times, and had only narrowly escaped total collapse in 1889. As we have seen, he was also involved as a founder member of the BAA itself, which had signed up 234 members even before its inaugural meeting, and only 11 months before he penned his letter to the *Cheshire Observer* the inaugural meeting of the BAA North-Western Branch had taken place at Manchester. There was clearly an appetite for astronomy abroad in the country and Longbottom was keen to cater to it.

The Chester Society for Natural Science, Literature and Art (sometimes referred to as the Chester Society of Natural Science & Art), had been operating from the City’s Grosvenor Museum (Figure 12) since its formation in 1871 by Charles Kingsley (1819 – 1875), Canon of Chester Cathedral (1870-1873) and author of The *Water Babies* and *Westward Ho!* (74). This satisfied some of the broad intellectual needs of Cestrians and encompassed many scientific pursuits such botany, zoology, entomology and microscopy, which were discussed in its *Proceedings* (Figure 13). Several members of the Society had an active interest in astronomy, including three members of the Dobie family who were well-known medical practitioners in the City. Dr William Murray Dobie (1828-1915) was a founder member of the Society and a noted lunar observer (75) and his sons, Dr. William Henry (1856-1946), one time a Chairman of the Zoological Section, and Dr. Herbert Murray (1865-1936), who coordinated the Society's interest in Lepidoptera. But there was no section dealing specifically with astronomy. So why, Longbottom reasoned, should Chester not have its own astronomical society?

Longbottom announced in the *Cheshire Observer* that a public meeting was to be held in Chester Town Hall on 21 November 1892 with the aim of establishing what he was already calling the "Chester Astronomical Society" (Figure 14), suggesting that: (73),

"any who 'consider the heavens', or feel even the remotest interest in the stars, cannot do better that unite in this endeavour to warm up what is sometimes rightly condemned as a cold and lonely cult"

He organised the meeting with another active amateur astronomer, Rev J. Cairns Mitchell, BD, FRAS, a BAA member who was minister at a Presbyterian Church in the City. It was chaired by Dr. Herbert Dobie and Cairns Mitchell informed the meeting that several interested people had recently agreed to establish a Chester Astronomical Society with Dr. Herbert Dobie as president and with Longbottom, W.E. Plummer of Bidston Observatory and himself as vice presidents. But things were destined not to run quite as smoothly as Longbottom and Cairns Mitchell had intended. Whether or not Longbottom and the others had considered forming an Astronomical Section within the existing Chester Society of Natural Science & Art, rather than a separate organisation, the move clearly upset the committee of the Society. As a result they sent a delegation to the public meeting with a counter-proposal that the new astronomical society should affiliate to the larger grouping. On the face of it this made sense: the Chester Society of Natural Science & Art was well established, well organised and it had access to the Grosvenor Museum for meetings. However, Longbottom explained that the main objection was the hefty membership fee that it levied, 5 shillings (25 pence), which might put people off, especially young people. By contrast the fee that he had in mind for the Chester Astronomical Society was half that. In the end, after much discussion, a solution was found: the new astronomical society would indeed affiliate to the established Society, with appropriate representation on its Committee, but with members paying the lower fee, although they would only have access to activities of the astronomy section (the committee of the Chester Society of Natural Science & Art felt sure that within a short time, seeing the benefits offered by the parent Society, most members of the astronomy branch would want to become full members of the Society in any case). The motion was proposed by Longbottom and carried. Thus in the end what could have been an awkward confrontation was resolved amicably to everyone's satisfaction. Whilst local astronomical societies of the modern era may be affected by political interventions of this sort from time to time, the events usually occur behind closed doors, rather than in the public gaze – by contrast, the events of the meeting were reported in great detail in the next edition of the *Cheshire Observer*! (76). It is of course possible that Longbottom had all along planned a public confrontation with the Society committee to ensure that the matter of the subscription could be dealt with head on.

After the eventful first meeting of what was now the "Astronomical Section of the Chester Society of Natural Science & Art", things calmed down. The first proper meeting of the Section was held the following month at the parent Society's headquarters, the Grosvenor Museum. A new section committee was voted in, with Cairns Mitchell as chairman and a committee comprising Longbottom, Dr. Herbert Dobie and three others. Following these business matters, Longbottom delivered a lecture on "Jupiter & Mars" (77).

The Astronomical Section continued to go from strength the strength as interest in the subject grew in the City. Perhaps this helped the Society to secure no less than three visits during the mid-1890s from one of the most celebrated and sought after public speakers on science of the age, Sir Robert Ball (1840-1913; Figure 15). Ball was Professor of Astronomy and Geometry at Cambridge University and towns and cities up and down the country vied with each other to obtain his services as speaker. As a youngster Ball had attended school in the village of Tarvin, just outside Chester, and the local connexion may therefore have had an influence. The subjects on which he spoke were: "Krakatoa, the mighty volcano" (November 1894), "Other worlds than ours" (January 1896) and "The Recent Eclipse" (October 1896). The last one was about the Norway eclipse, but because it had been clouded out the local paper noted that Ball needed to rely on his famous humour and repartee more than usual to keep the audience interested and amused! (78) At the end of the meeting Ball publically thanked Longbottom, who had of course been a member of the eclipse expedition, for allowing him to show some of his photographs during the talk.

Longbottom's clash with the Chester Society of Natural Science & Art was soon forgotten. He gave several talks (79) at its main meetings and he became involved in many of its activities outside of astronomy. Figure 16 shows him on a Society excursion to Burton Point Iron Age Hill Fort in Cheshire. He went on the serve as the Society's president during the 1906-1907 session and as vice president 1911-1912. In 1906 he was awarded its Kingsley Memorial Medal, named after Charles Kingsley, for his work on astronomy. The Medal was awarded annually to a resident of the City who had contributed materially to the advancement of science. It was presented at Chester Town Hall by the Duchess of Westminster (80). A public exhibition of the Society's work was staged at the same time in which Longbottom exhibited his equatorial mount fitted for astronomical photography and some photographs of the 1905 solar eclipse taken by himself and by Sir William Christie (1845-1922), the Astronomer Royal.

**Longbottom's astronomical legacy in Chester**

As we have already seen, Longbottom left Chester in the early 1920's.  The Astronomical Section of the Chester Society of Natural Science & Art continued through the inter-war years. However, by the time the Second World War came to an end the Section was effectively dormant. In 1950 Richard Baum, a well-known BAA member who still lives in Chester, was instrumental in revitalising the Section, along with Paul Taylor who was a student at the Chester Teacher Training College, serving as Chairman and Secretary, respectively (81). The group thrived well into the 1960s. In November 1969, following the excitement surrounding the Apollo Moon landings and a number of articles in the *Chester Chronicle* on astronomy, there appeared to be a resurgence in interest in astronomy and space amongst Cestrians. As a result Richard Baum suggested forming a group other than the Astronomy Section of the Chester Society of Natural Science & Art. Thus the first meeting of the new society was held in January 1970. It was named the Chester Astronomical Society (CAS), the same name as Longbottom's originally proposed organisation of 1892. The events leading to the formation of the CAS will be discussed in more detail a future paper.

Thus some 120 years after Longbottom helped to establish the Astronomical Section of the Chester Society of Natural Science & Art, its descendent, the Chester Astronomical Society, continues to thrive to this day, holding monthly meetings during which a range of professional and amateur astronomers bring the latest developments in astronomy to Cestrians. It is perhaps a fitting tribute to one who had so much enthusiasm for astronomy and who was instrumental in encouraging and organising amateur astronomy at a local and a national level: Frederick William Longbottom.

**Acknowledgements**

I am most grateful for the assistance I have received from a great many people whilst preparing this paper. Richard Baum has been an invaluable source of encouragement, as well as providing his personal insight into Chester's astronomical history. He also commented on an early draft of the paper. Kate Riddington, Keeper of Natural History at the Grosvenor Museum as Chester, answered many enquiries and hosted my visit to the Museum to inspect a number of publications concerning Longbottom. She also provided copies of his portrait and the photograph of the Members of the Chester Society of Natural Science & Art at Burton Point Hill Fort. Dr. Richard McKim kindly prepared Longbottom's photograph of Venus, which resides in the BAA Mercury & Venus Section archive, for publication. Peter Hingley, RAS Librarian, provided information about the Reynolds telescope at Helwân. Tony Kinder researched Longbottom's Census returns. Mark Butterworth allowed me to use a copy of Longbottom's photograph of the moon.

This research made extensive use of scanned back numbers of the *Journal*, which exist largely thanks to the efforts of Sheridan Williams, as well as of the *English Mechanic*, thanks to Eric Hutton. These are truly wonderful resources for historians of nineteenth and twentieth century British astronomy. I also used the NASA/Smithsonian Astrophysics Data System and the British Newspaper Archive (British Library).

**Address:** “Pemberton”, School Lane, Bunbury, Tarporley, Cheshire, CW6 9NR, UK

BAA Handbook. After Tattenhall, he moved to the Wirral and then spent the last few years of his life in Chester itself.

54. Harry Krauss Nield lived in Sale, Cheshire. He also viewed the 1905 eclipse in Spain. He became FRAS in 1902. In later years he stood as a Parliamentary candidate for the Liberal party first at St. Albans, Herts., then at Macclesfield, Cheshire. He died at Barnet, Herts., on 26 October 1926. His obituary appears in MNRAS, 87, 259 (1927).

55. Levander F.W., Mem. Brit. Astron. Assoc. (1905).

56. Further details about the Brooks may be found in: Shears J., JBAA, 122, 17-30 (2012).

57. He showed these at a meeting of the Chester Natural Science Society: JBAA, 17, 48 (1906).

58. Observers' impressions of the eclipse, including those of Longbottom, were discussed at the BAA meeting on 26 October 1927: JBAA, 38, 1-20 (1927). Unfortunately his observing location is not stated.

59. He gave a talk of his Egyptian sojourn at a meeting of the Liverpool Astronomical Society entitled “An Astronomical Holiday” in 1912: see JBAA, 24, 131 (1913).

60. Reynolds described the Helwân Observatory and the telescope construction project at a meeting of the RAS in May 1907: JBAA, 17, 320 (1907). Reynolds himself had a 24 inch reflector at his Observatory at Harborne, near Birmingham, UK. An account of the history of the observatory can be read in Fesenkov V.G., Soviet Astronomy, 2, 256-259 (1958).

61. Writing in 1909, following a visit to Helwân, H.H. Turner had encouraged astronomers to visit: "I have no hesitation in advising others to make the most of any pretext for paying a visit", Turner H.H., Obs., 32, 111 (1909).

62. Longbottom describes developing plates at Helwân in JBAA, 34, 170 (1924). Several plates prepared by Longbottom were used to illustrate "Observations of Nebulæ Made During 1909-1911", Know-Shaw H., HelOB, 9, 69-78 (1912).

63. Knox-Shaw was the first person to photograph Comet Halley on its return to the 1910 perihelion, using the Reynolds telescope on 24 August 1909. He also photographed many other comets, including several during the winter of 1911/12, during Longbottom's secondment. It is possible that Longbottom assisted, given his interest in comet photography. In 1913, Knox-Shaw succeeded B.F.E. Keeling (1880-1919) who had been director at Helwân since 1905.

64. The author has written a paper on Knox-Shaw and the Helwân Observatory with Dr. Ashraf Shaker: Shears J. & Shaker A., JBAA submitted (2012).

65. Knox-Shaw H., HelOB, 15, 129-138 (1915).

66. Longbottom and Newstead's survey of the shore at Meols was described as "the last recorded antiquarian visit to Meols in the nineteenth century tradition": Griffiths D., Philpott R.A. & Egan G., "Meols: The archaeology of the north Wirral coast", Oxford. University Institute of Archaeology Monograph 68, publ. University of Oxford (2007).

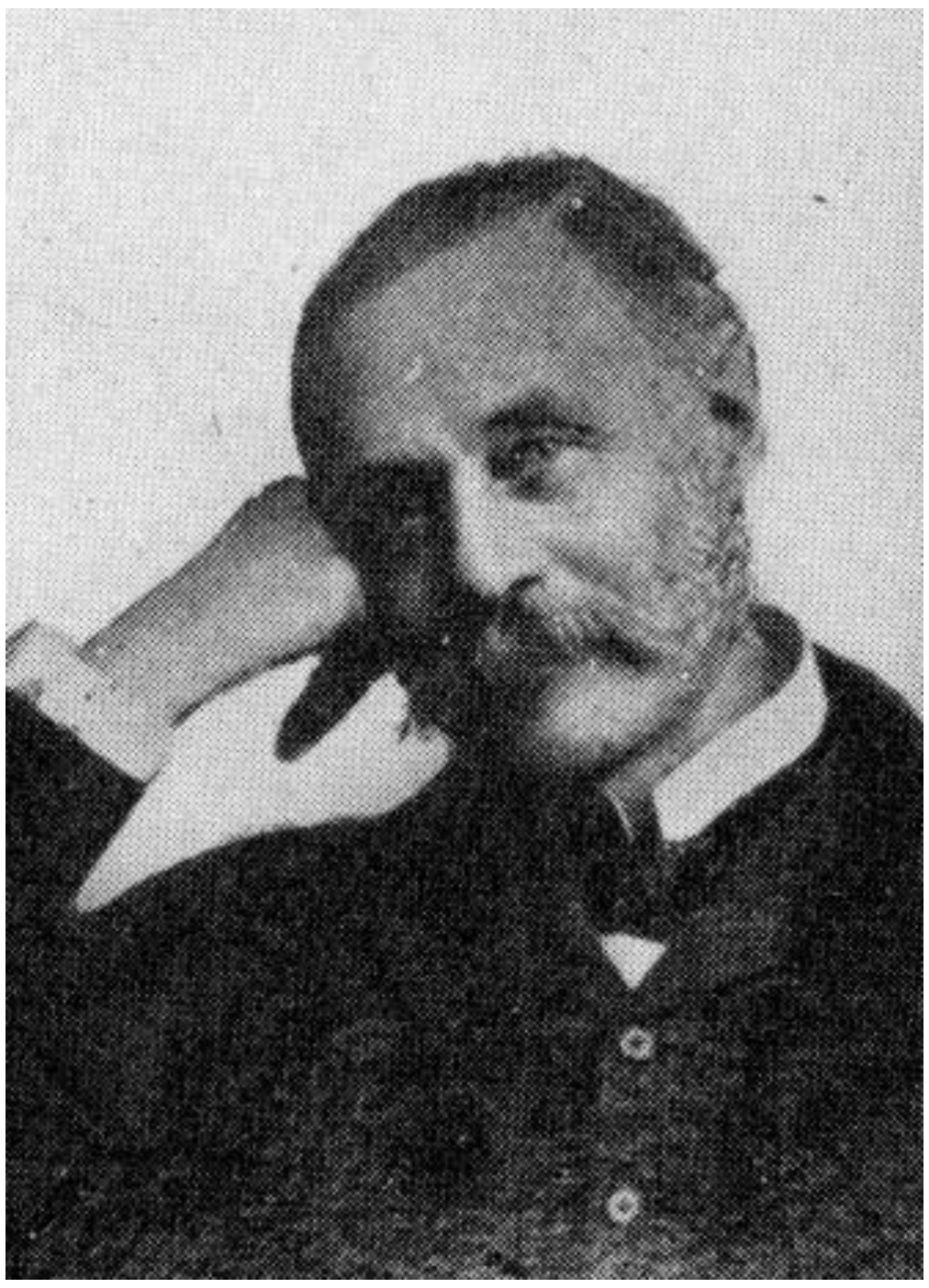

Figure 1: Frederick William Longbottom (1850-1933) (82)

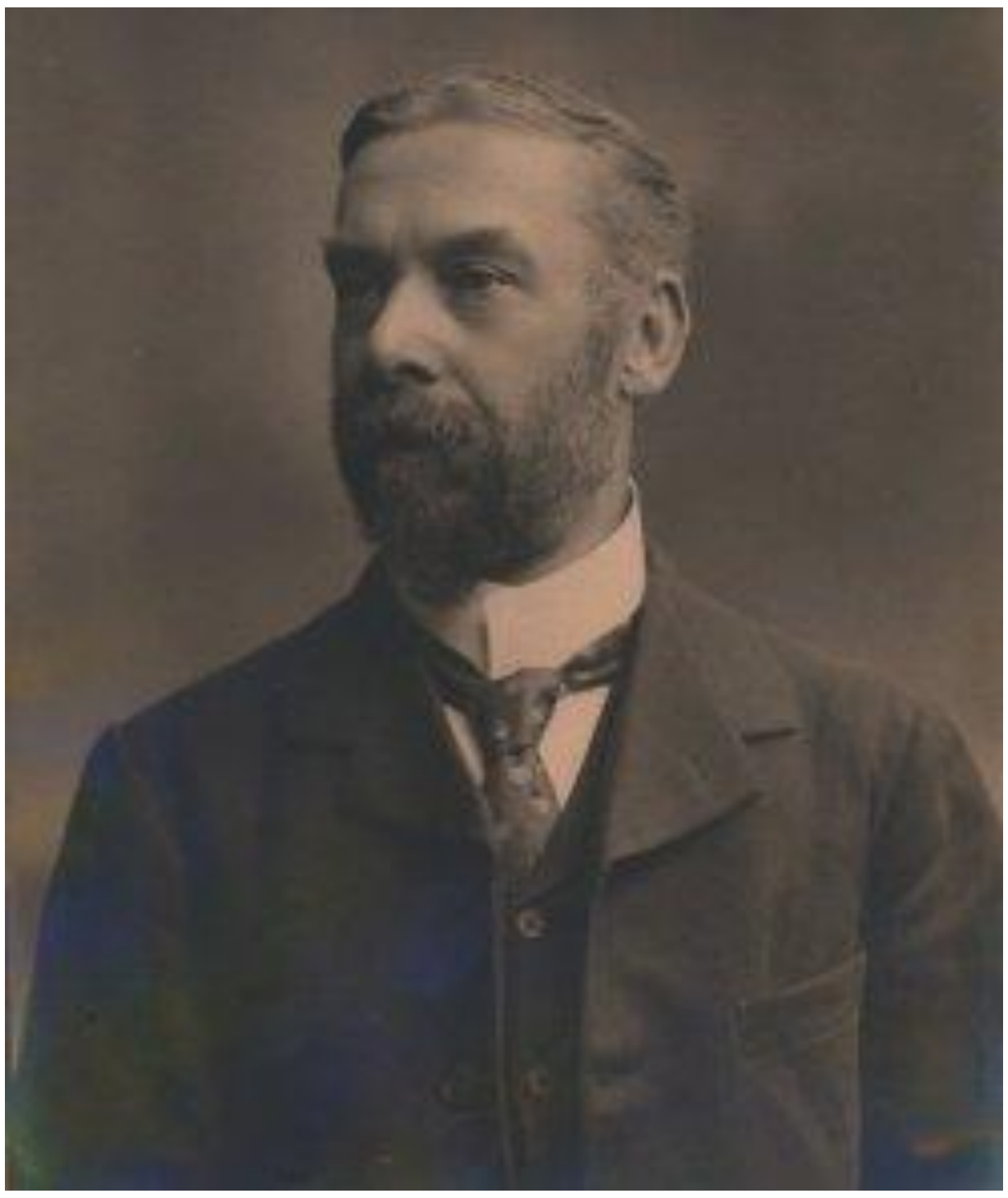

Figure 2: W.E. Plummer (1849-1928), Director of the Liverpool Observatory at Bidston Hill

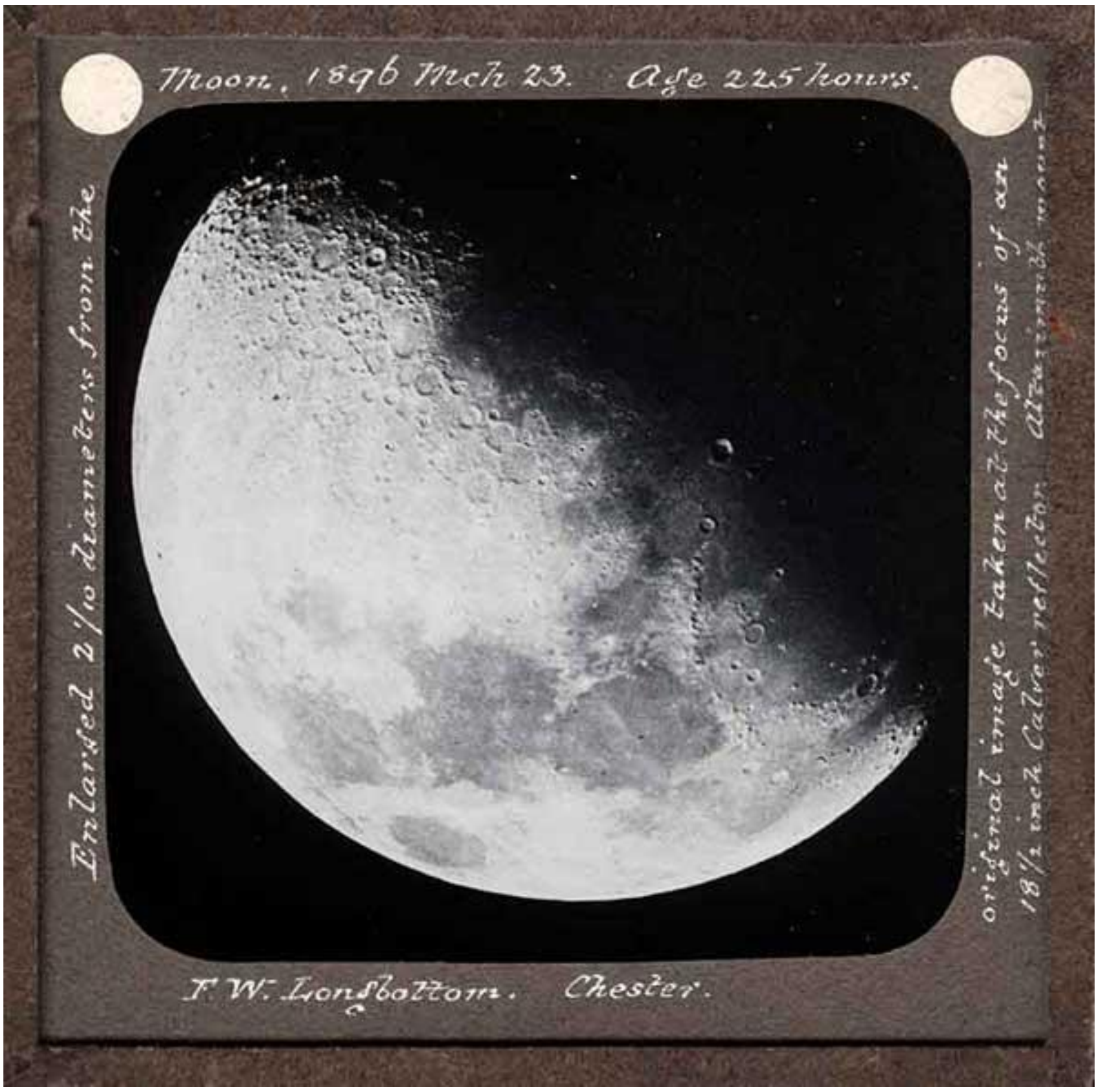

Figure 3: Photograph of the Moon by F.W. Longbottom on 23 March 1896 with his 18½ inch Calver from Chester. This lantern slide is a 2.1 times enlargement by Longbottom of his original exposure.

(Image courtesy of Mark Butterworth)

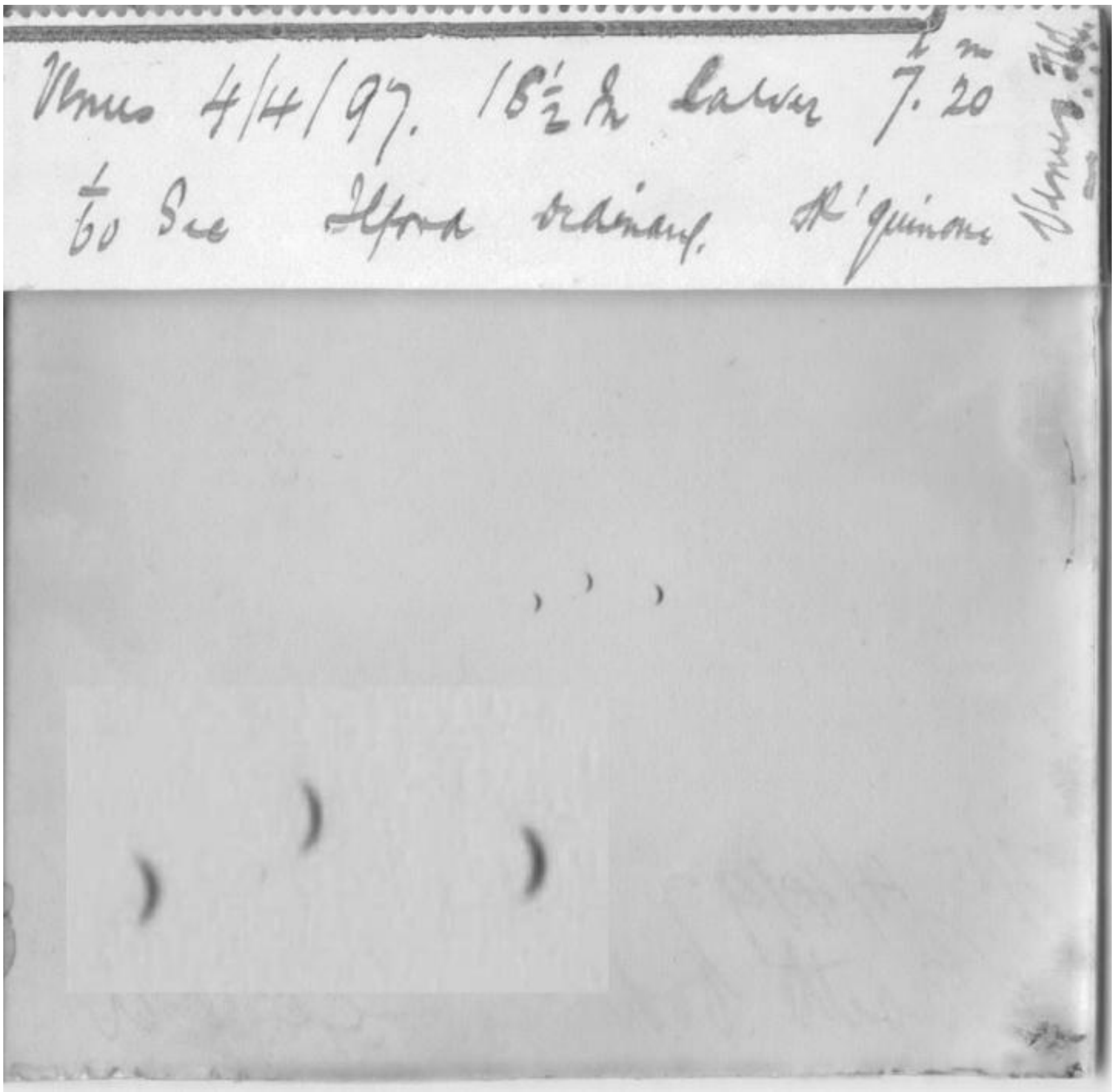

Figure 4: Photographs of Venus by F.W. Longbottom on 4 April 1897 at 7.20 pm. 1/60th second exposure with the same telescope as in Figure 4. The original plate is 81 mm square. The actual images are the small ones – the inset shows a x4 expanded view.

(Image of original plate courtesy of Dr. Richard McKim)

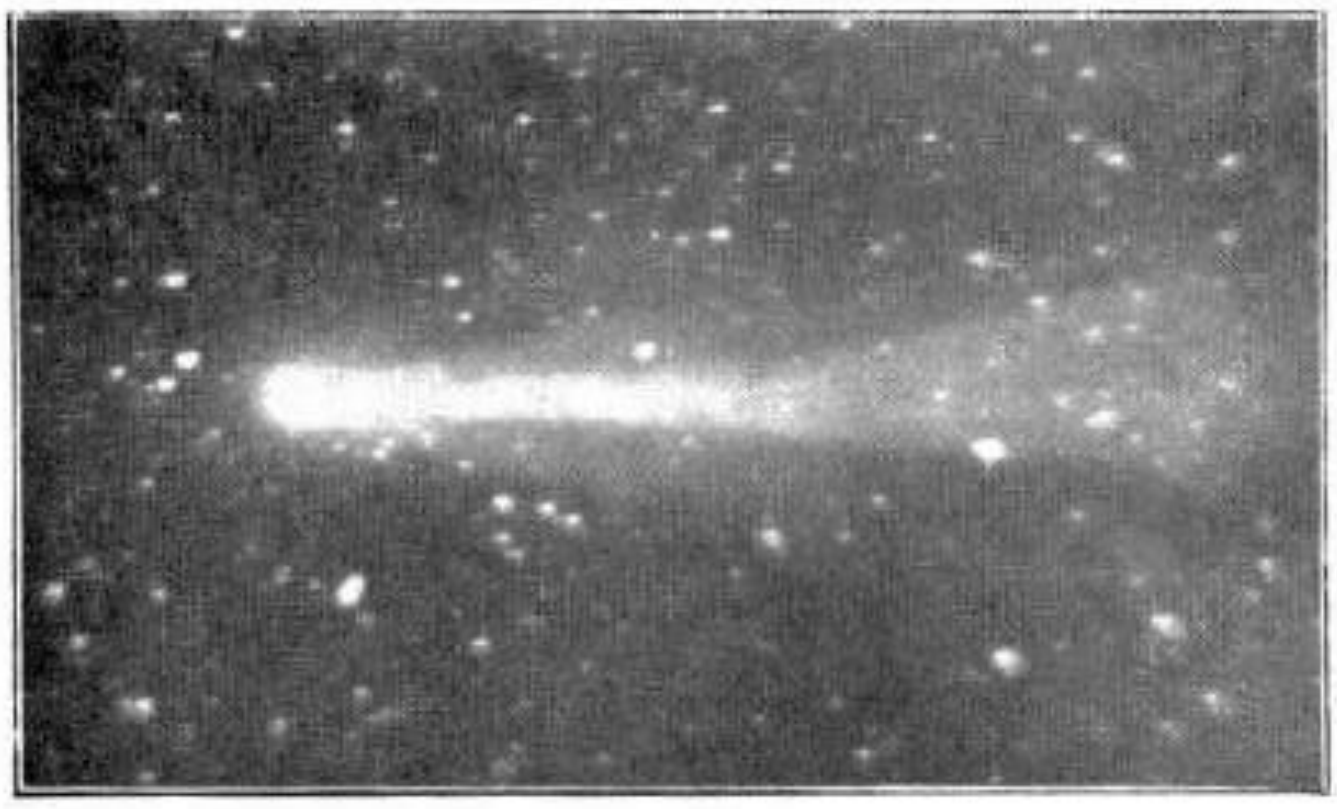

October 12. 10 minutes' exposure.

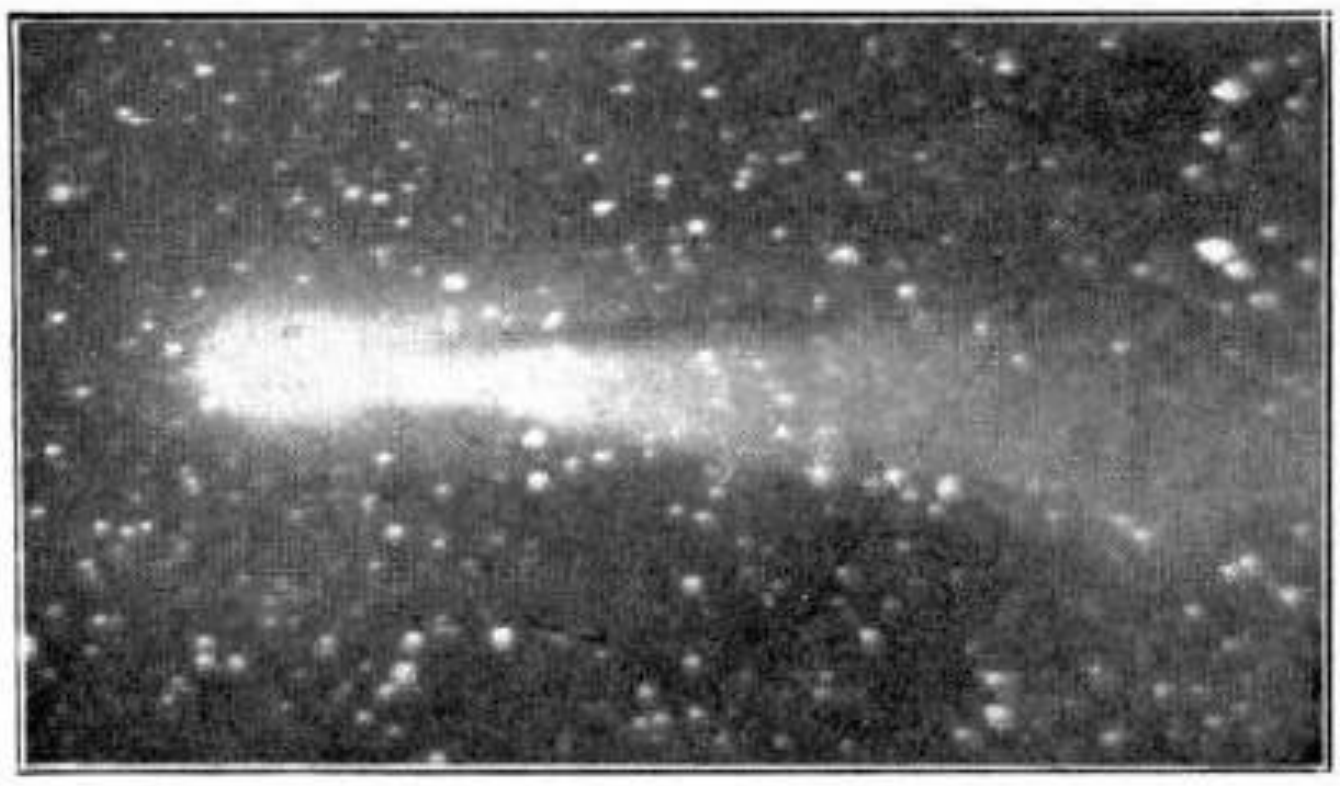

October 13 20 minutes' exposure.

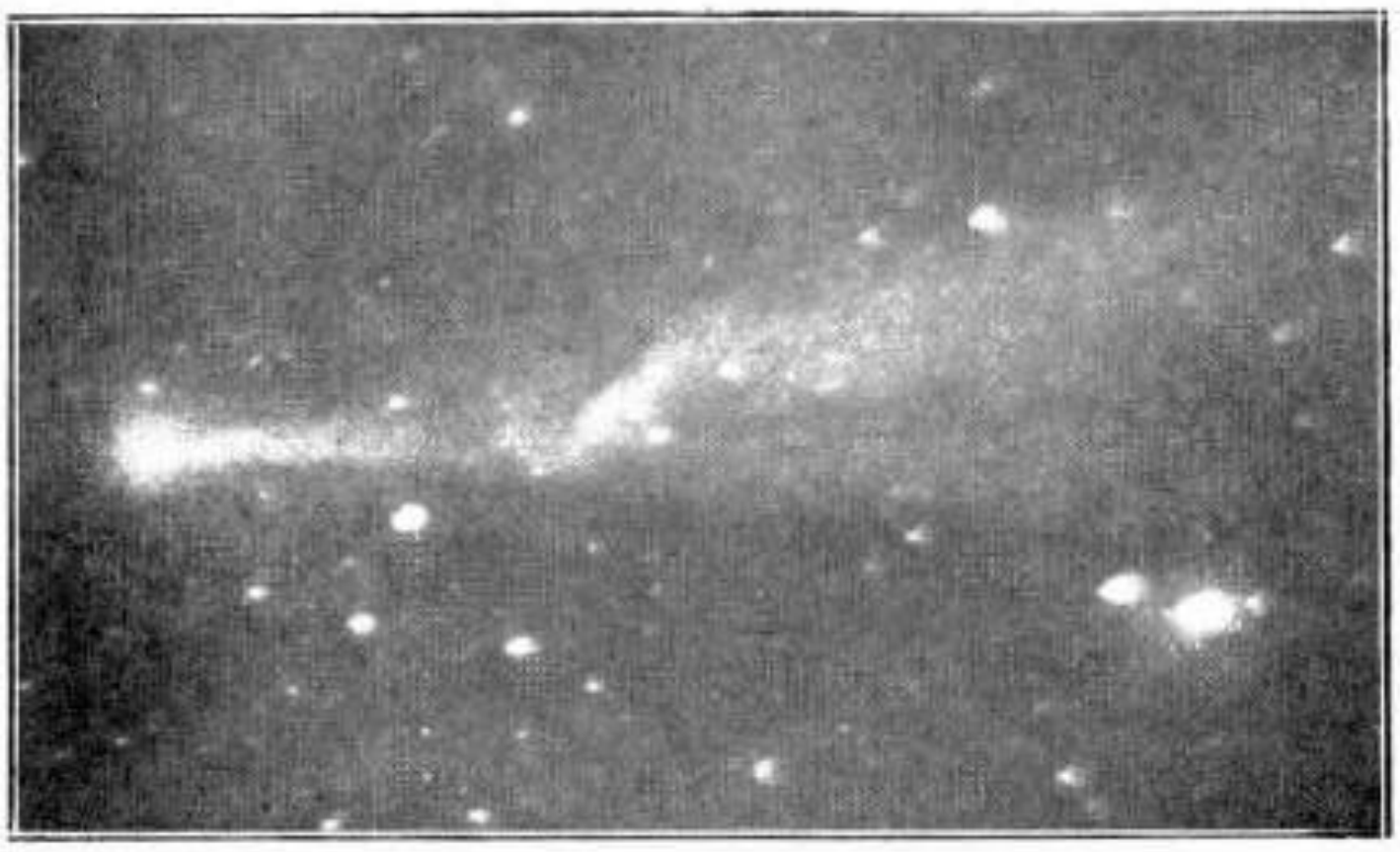

October 15. 5 minutes' exposure.

Figure 5: Comet Morehouse 1908 R1. Longbottom used his 12½ inch f/1.9 reflector stopped down to 9 inches (f2.7) with Ilford "Monarch" plates (36)

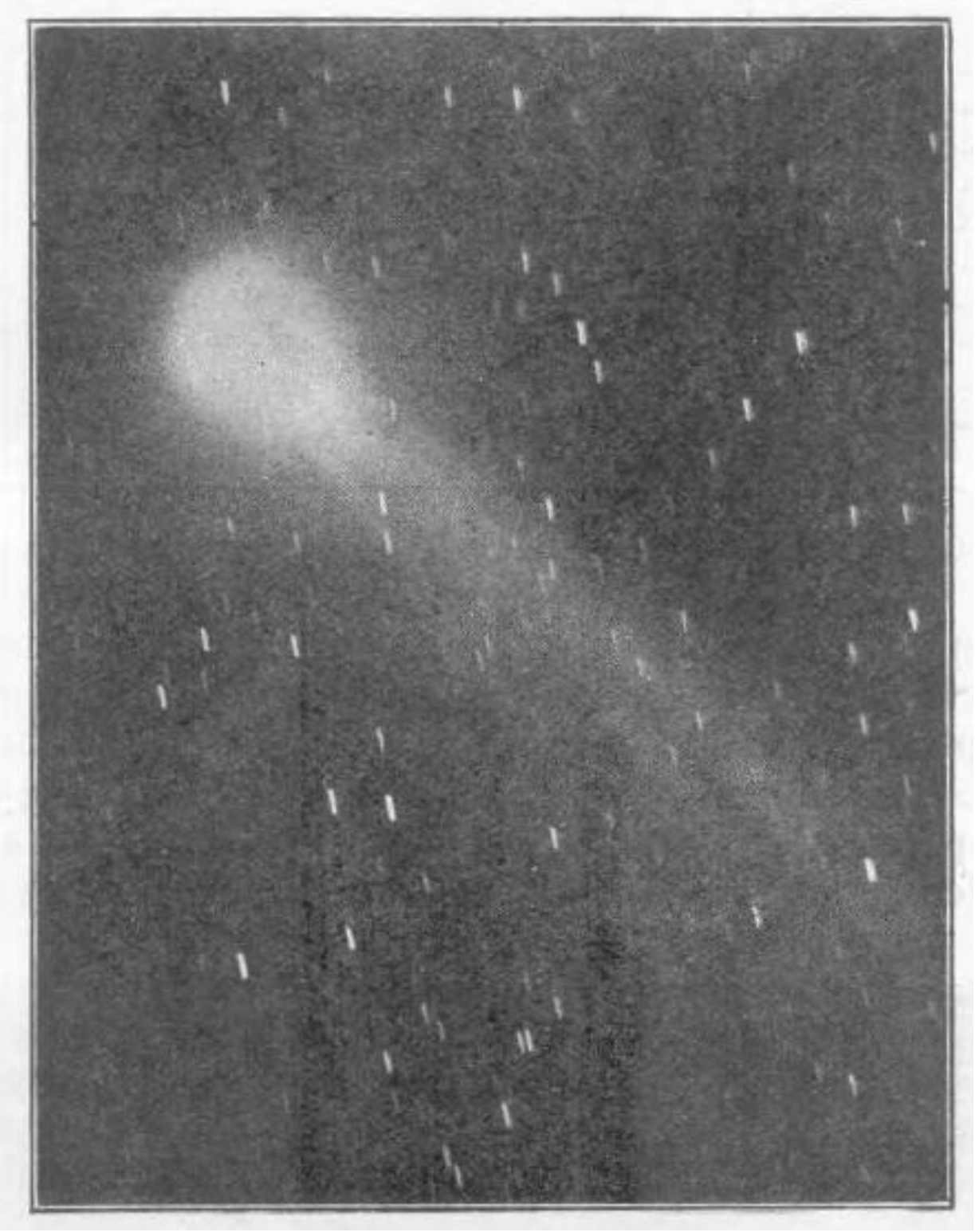

(a)

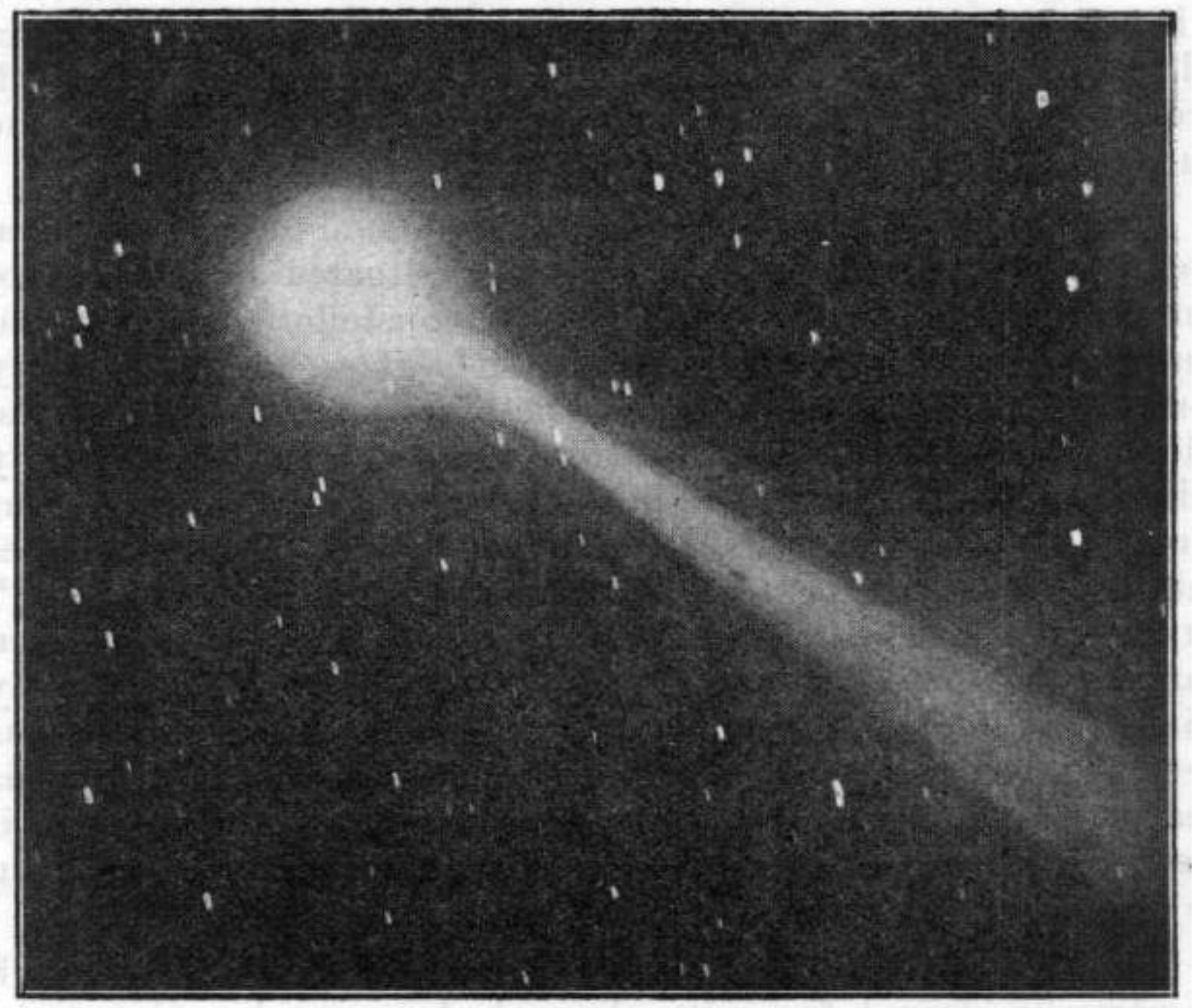

(b)

Figure 6: Comet Brooks 1911 O1 photographed by Longbottom with his 15-in. f/4 Cope reflector (a) 21 September 1911, 30 min exposure (83) (b) 24 September 1911, exposure not stated (37)

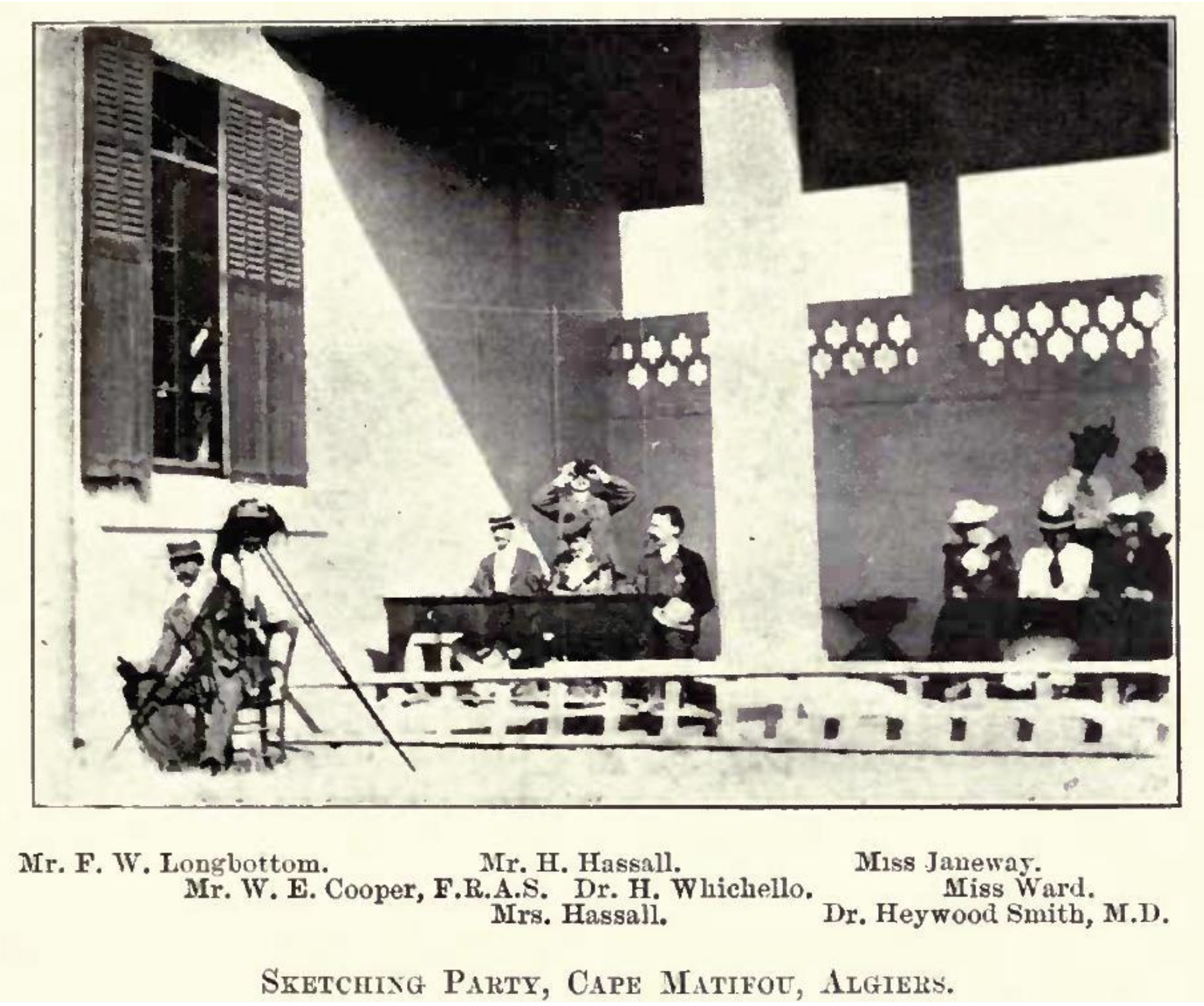

Figure 7: Observers at the 1900 solar eclipse in Algiers. Longbottom can be seen at left with his camera. His friend from Chester, Dr. Harold Whichello, is seated at the desk

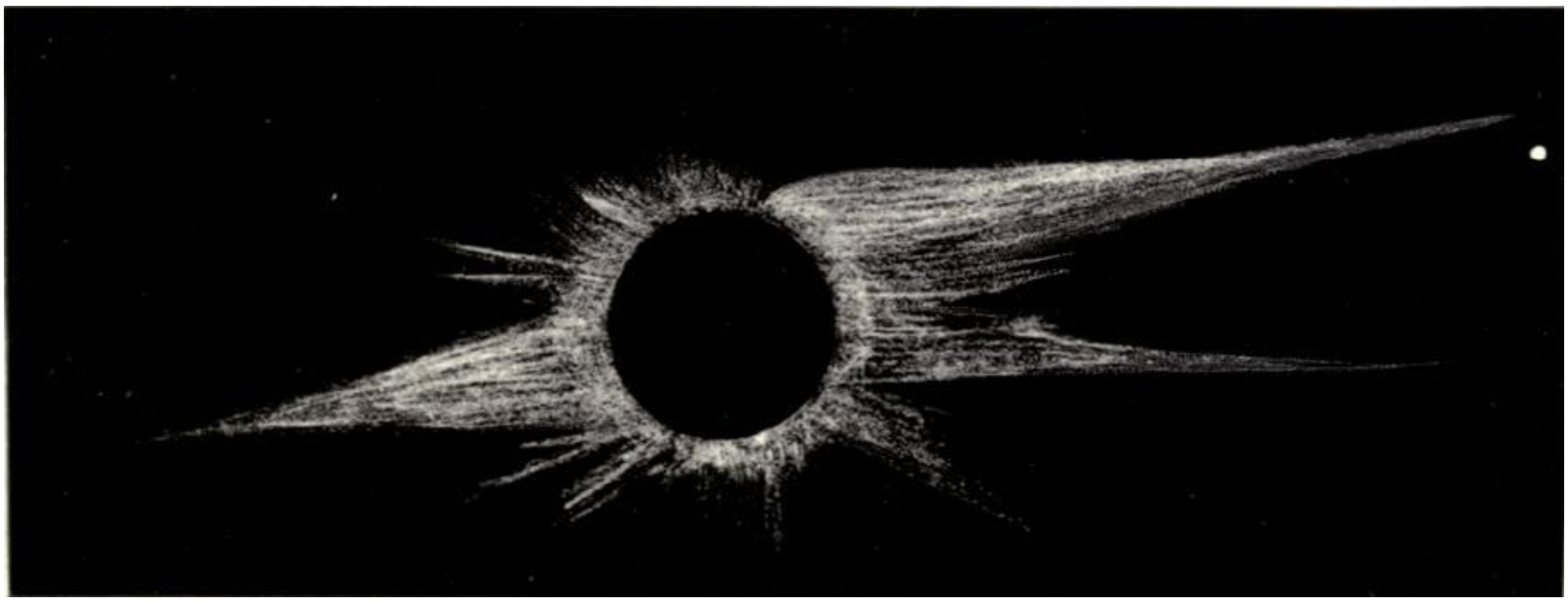

Figure 8: Sketch by H. Krauss Nield of the corona during the 1900 eclipse observed at Algiers

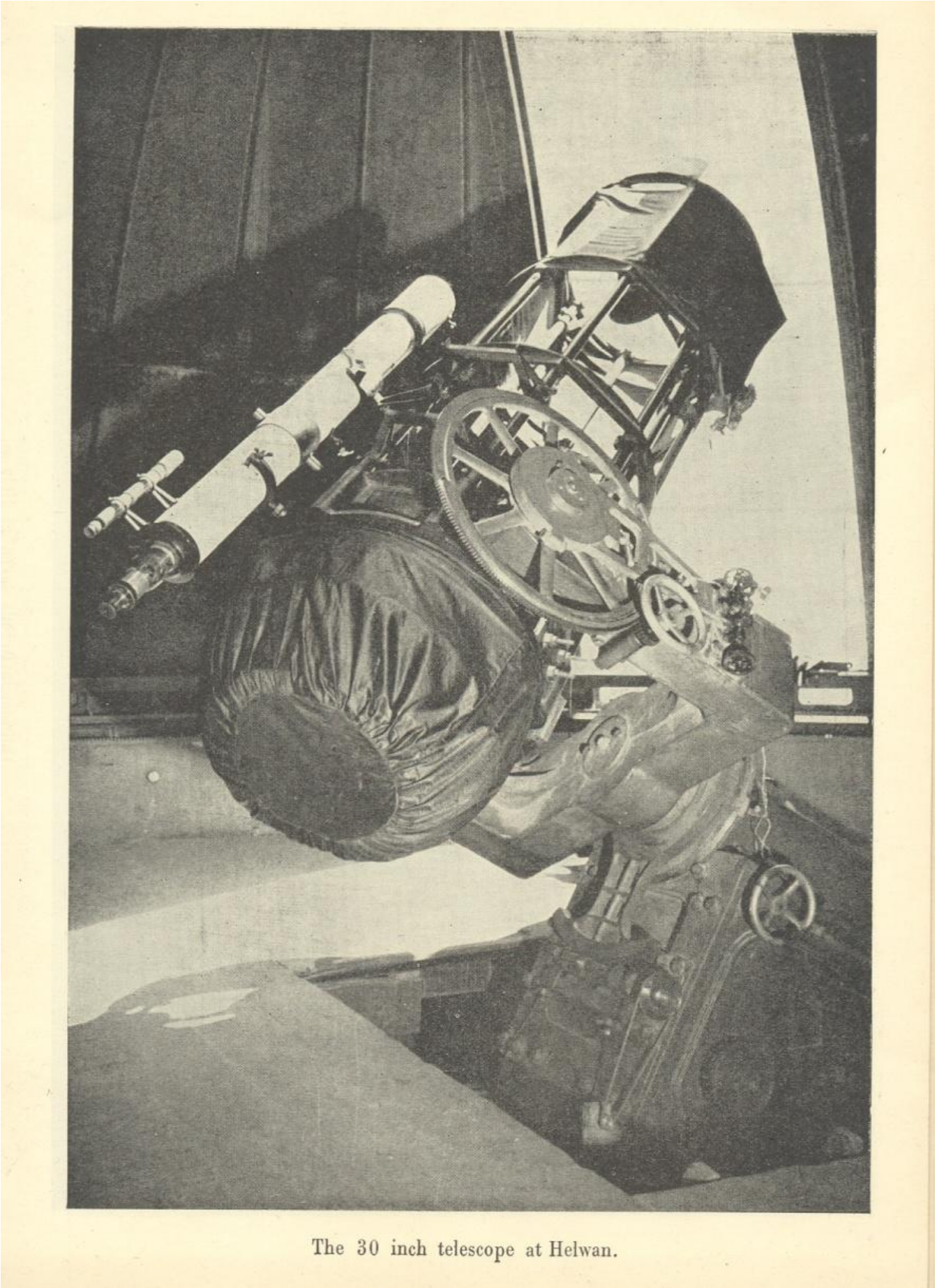

Figure 9: The 30 inch Reynolds reflector at Helwân Observatory.

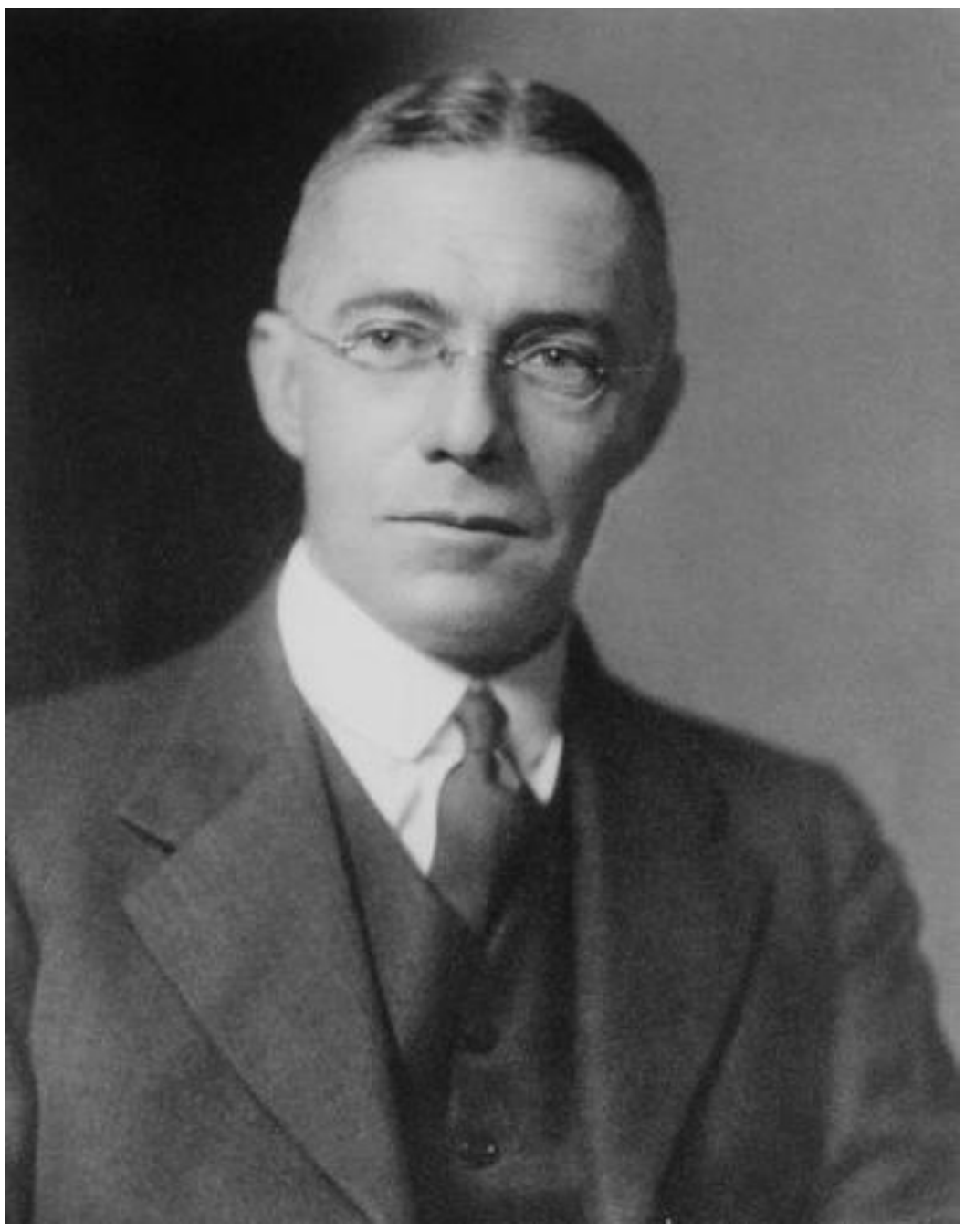

Figure 10: Harold Knox-Shaw (1885-1970)

(Royal Astronomical Society)

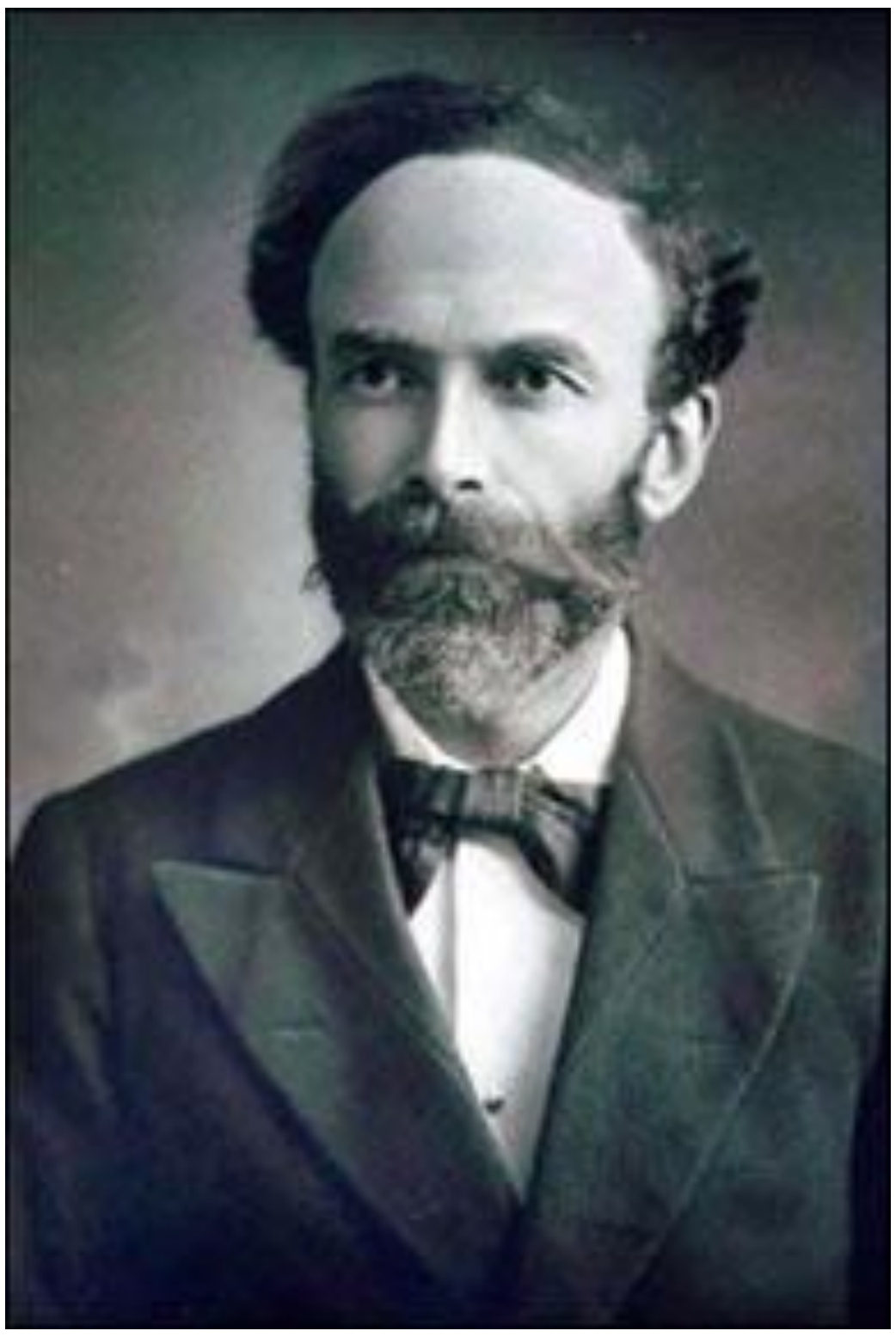

Figure 11: Prof. Robert Newstead, FRS (1859 – 1947)

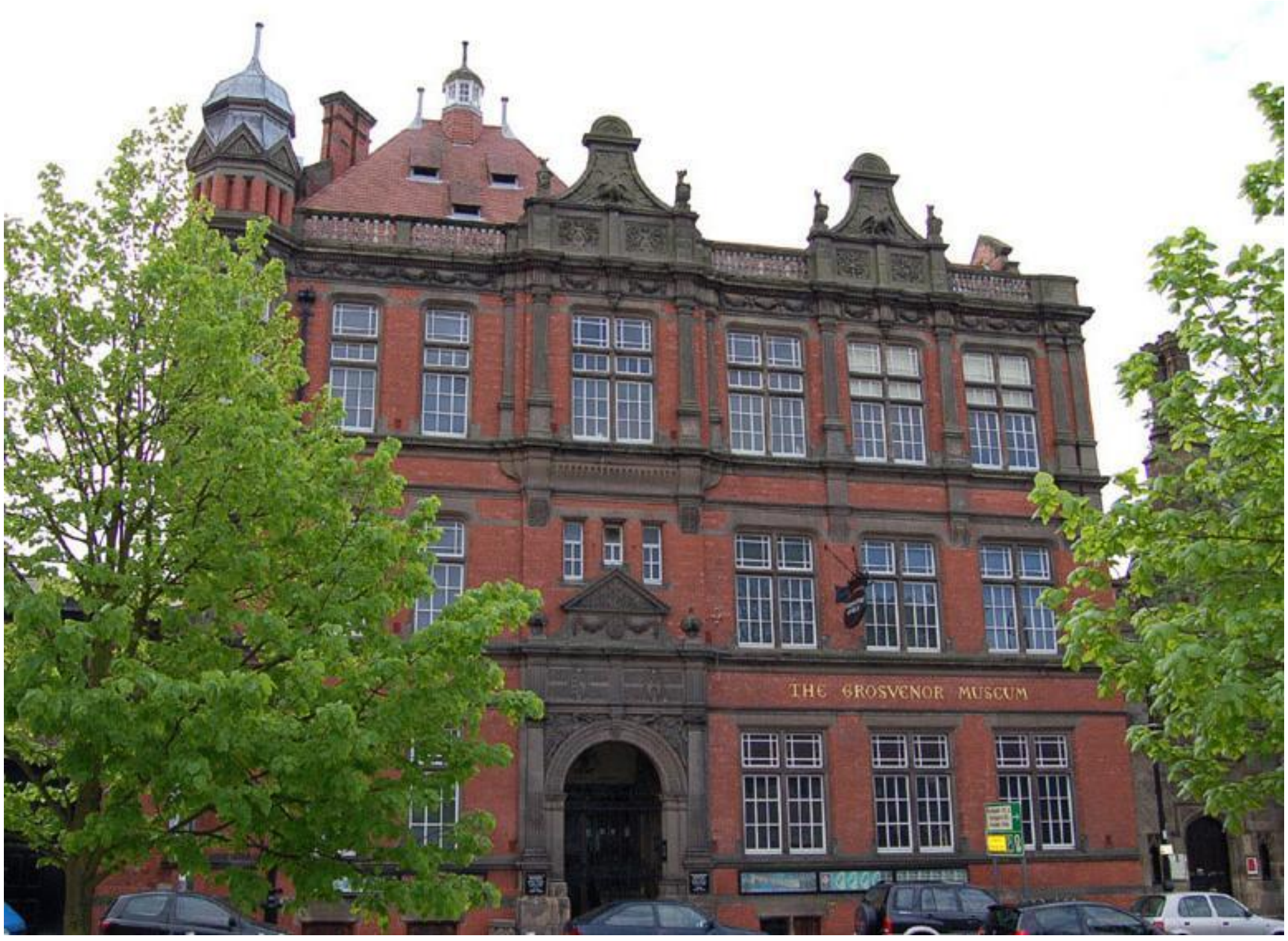

Figure 12: The Grosvenor Museum, Chester

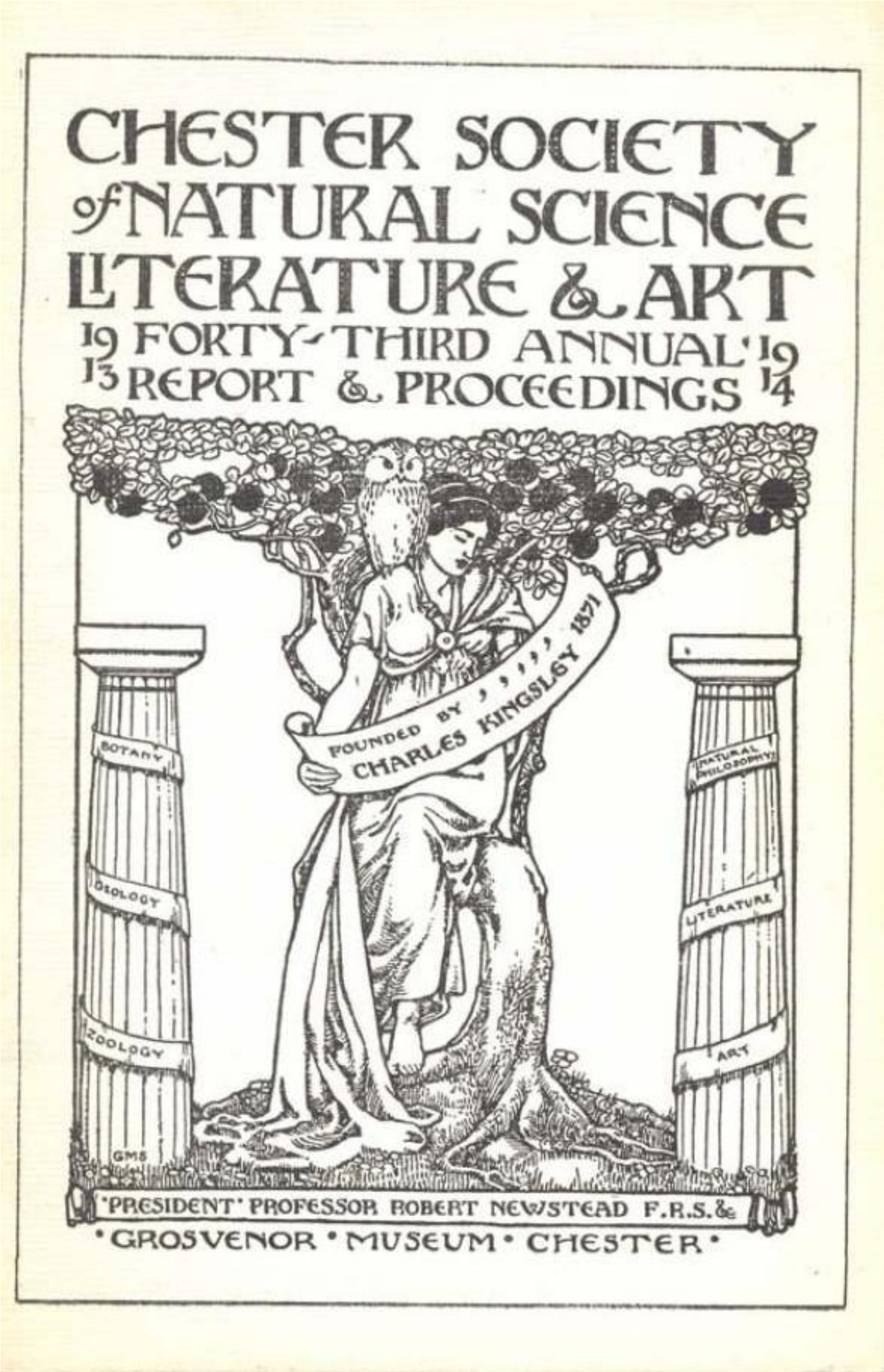

Figure 13: Frontispiece of the 1913-4 *Report & Proceedings of The Chester Society for Natural Science, Literature and Art*

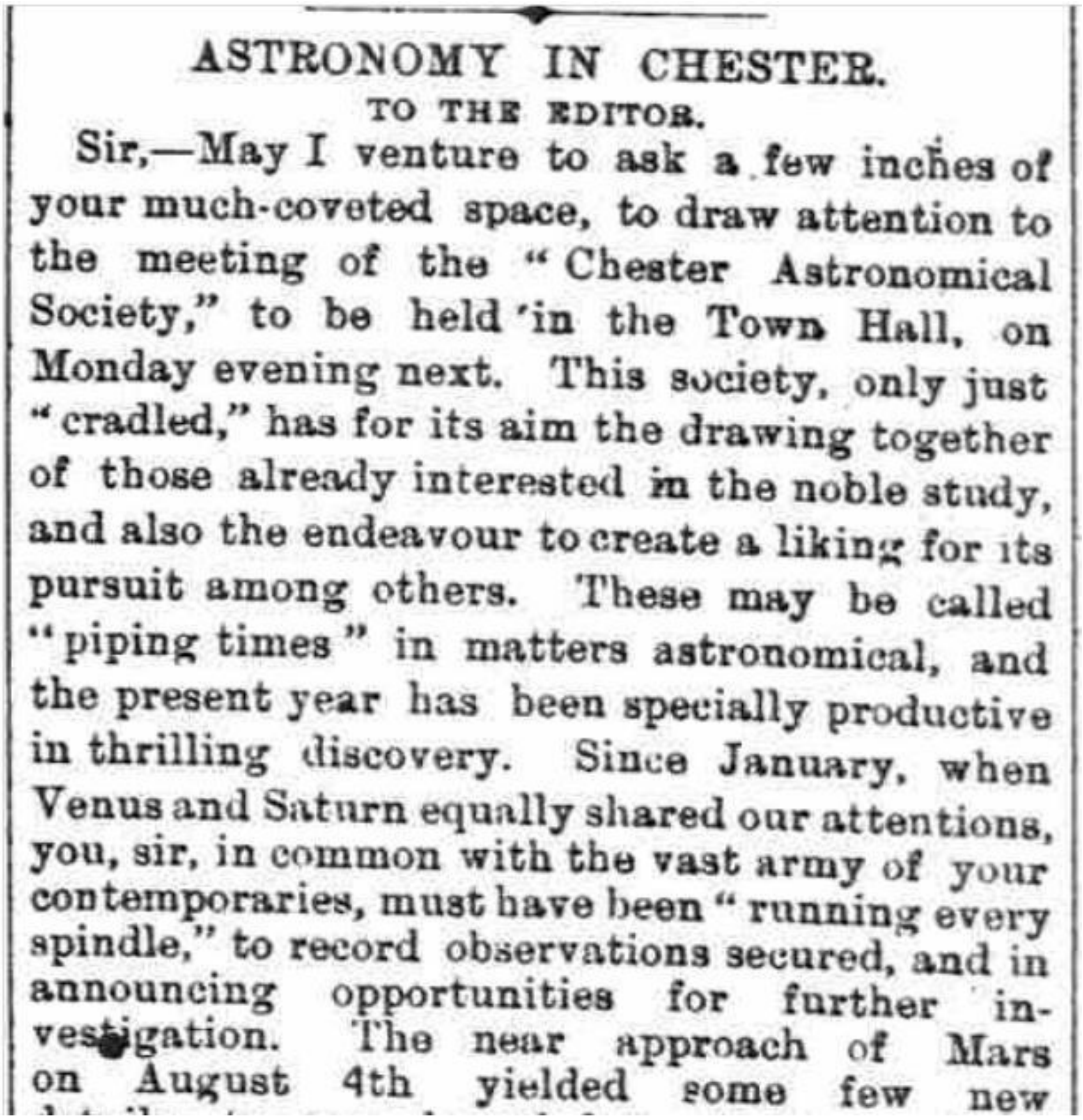

ASTRONOMY IN CHESTER.

TO THE EDITOR.

Sir,—May I venture to ask a few inches of your much-coveted space, to draw attention to the meeting of the "Chester Astronomical Society," to be held in the Town Hall, on Monday evening next. This society, only just "cradled," has for its aim the drawing together of those already interested in the noble study, and also the endeavour to create a liking for its pursuit among others. These may be called "piping times" in matters astronomical, and the present year has been specially productive in thrilling discovery. Since January, when Venus and Saturn equally shared our attentions, you, sir, in common with the vast army of your contemporaries, must have been "running every spindle," to record observations secured, and in announcing opportunities for further investigation. The near approach of Mars on August 4th yielded some few new

Figure 14: Part of Longbottom's letter announcing the formation of the proposed "Chester Astronomical Society". *Cheshire Observer* 19 November 1892

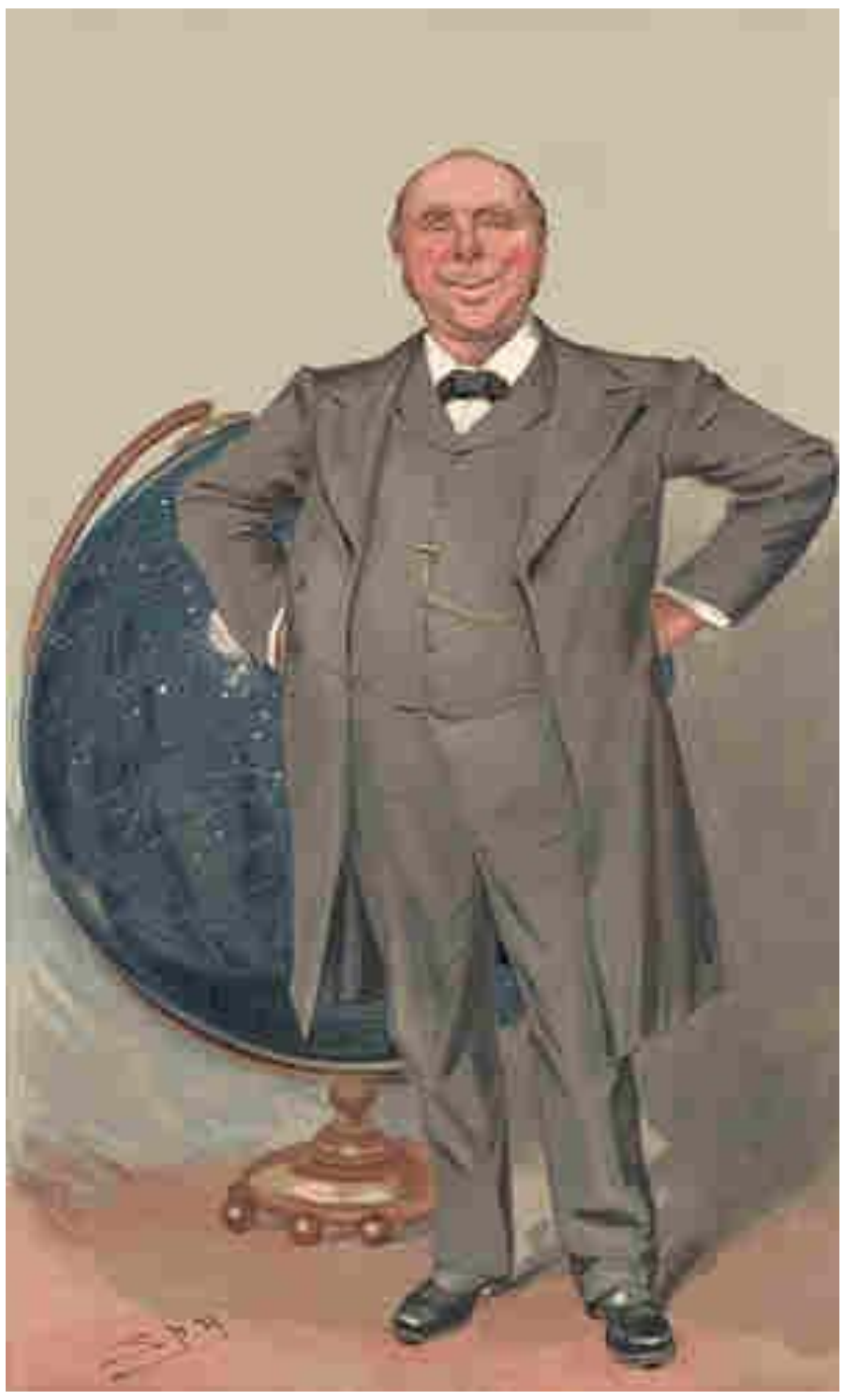

Figure 15: Sir Robert Stawell Ball (1840-1913), *Vanity Fair* print by Leslie Ward (1851-1922)

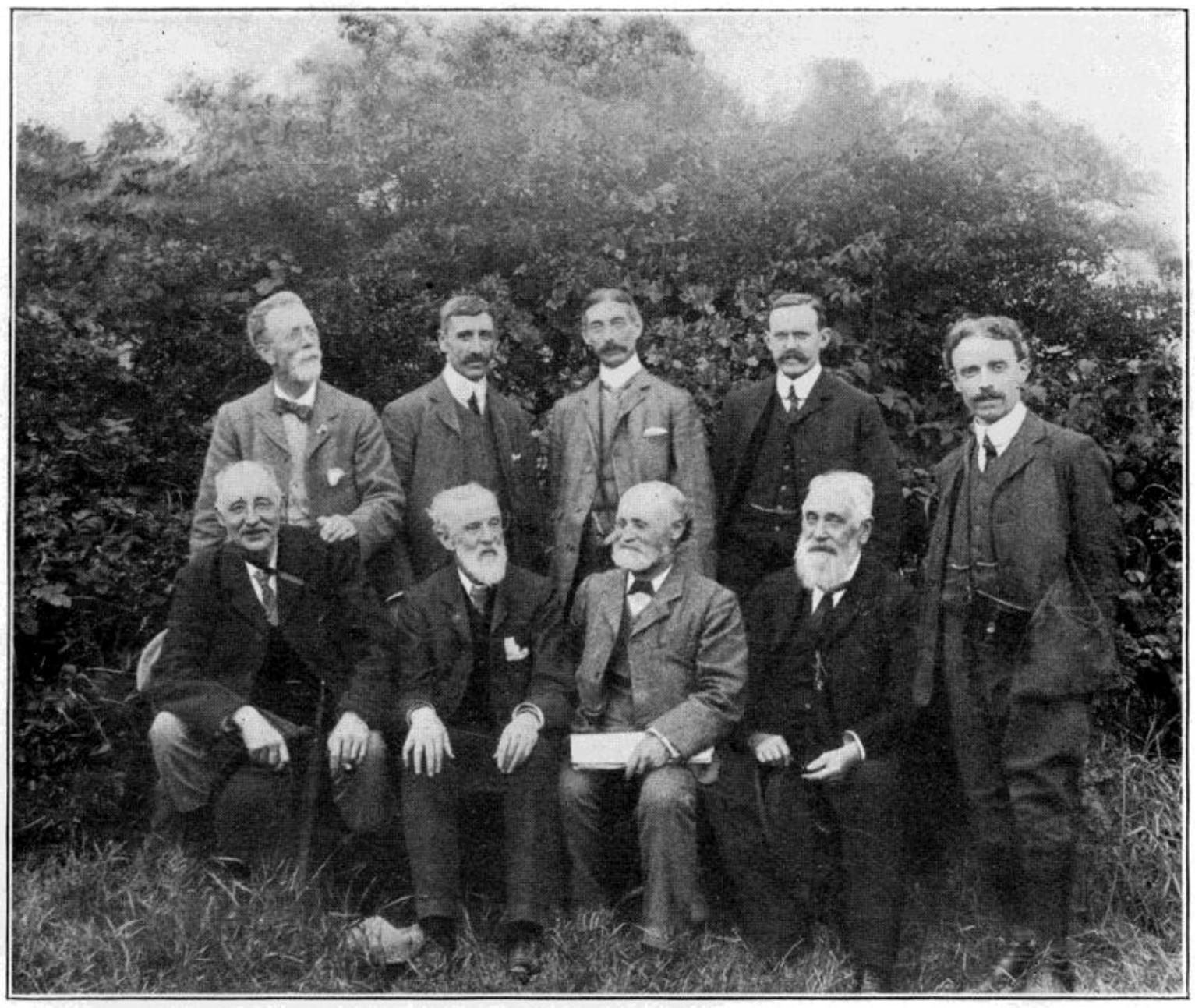

At Burton Point, 1907

J. D. Siddall; F. Simpson; F. W. Longbottom; W. F. J. Shepheard; A. Newstead;
W. Shone; Thos. Shepheard; A. W. Lucas; H. Stolterfoth

Figure 16: Members of the Chester Society of Natural Science & Art at Burton Point Hill Fort in Cheshire, 1907 (82). Longbottom is in the middle of the back row.